\documentclass[8.5pt,twoside]{article}
\usepackage{color}
\usepackage{float}
\usepackage{units}
\usepackage{amsmath}
\usepackage{amssymb}
\usepackage{wasysym}
\usepackage{graphicx}
\usepackage{esint}
\usepackage{appendix}

\usepackage[super,sort&compress,comma]{natbib} 
\usepackage{times,mathptm}
\usepackage{sectsty}
\usepackage{balance} 
\usepackage[text={11.3cm,20.4cm},centering]{geometry} 

\usepackage{graphicx} 
\usepackage{lastpage}
\usepackage[format=plain,justification=raggedright,singlelinecheck=false,font=small,labelfont=bf,labelsep=space]{caption} 
\usepackage{fancyhdr}
\usepackage{amsmath}
\usepackage{amssymb}
\usepackage{wasysym}
\usepackage{graphicx}
\usepackage{esint}
\usepackage{appendix}

\begin{document}

\thispagestyle{plain}
\fancypagestyle{plain}{
\fancyhead[L]{\includegraphics[height=8pt]{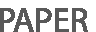}}
\fancyhead[C]{\hspace{-1cm}\includegraphics[height=15pt]{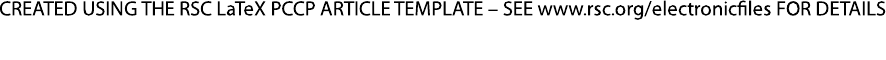}}
\fancyhead[R]{\includegraphics[height=10pt]{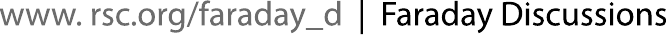}\vspace{-0.2cm}}
\renewcommand{\headrulewidth}{1pt}}
\renewcommand{\thefootnote}{\fnsymbol{footnote}}
\renewcommand\footnoterule{\vspace*{1pt}%
\hrule width 11.3cm height 0.4pt \vspace*{5pt}} 
\setcounter{secnumdepth}{5}

\makeatletter 
\renewcommand{\fnum@figure}{\textbf{Fig.~\thefigure~~}}
\def\subsubsection{\@startsection{subsubsection}{3}{10pt}{-1.25ex plus -1ex minus -.1ex}{0ex plus 0ex}{\normalsize\bf}} 
\def\paragraph{\@startsection{paragraph}{4}{10pt}{-1.25ex plus -1ex minus -.1ex}{0ex plus 0ex}{\normalsize\textit}} 
\renewcommand\@biblabel[1]{#1}            
\renewcommand\@makefntext[1]%
{\noindent\makebox[0pt][r]{\@thefnmark\,}#1}
\makeatother 
\sectionfont{\large}
\subsectionfont{\normalsize} 

\fancyfoot{}
\fancyfoot[LO,RE]{\vspace{-7pt}\includegraphics[height=8pt]{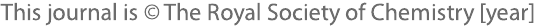}}
\fancyfoot[CO]{\vspace{-7pt}\hspace{5.9cm}\includegraphics[height=7pt]{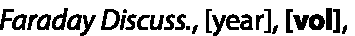}}
\fancyfoot[CE]{\vspace{-6.6pt}\hspace{-7.2cm}\includegraphics[height=7pt]{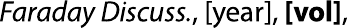}}
\fancyfoot[RO]{\scriptsize{\sffamily{1--\pageref{LastPage} ~\textbar  \hspace{2pt}\thepage}}}
\fancyfoot[LE]{\scriptsize{\sffamily{\thepage~\textbar\hspace{3.3cm} 1--\pageref{LastPage}}}}
\fancyhead{}
\renewcommand{\headrulewidth}{1pt} 
\renewcommand{\footrulewidth}{1pt}
\setlength{\arrayrulewidth}{1pt}
\setlength{\columnsep}{6.5mm}
\setlength\bibsep{1pt}

\noindent\LARGE{\textbf{A mesoscopic model for the rheology of soft amorphous solids, with
application to microchannel flows}}
\vspace{0.6cm}

\noindent\large{\textbf{Alexandre Nicolas,\textit{$^{a}$}and Jean-Louis Barrat,$^{\ast}$\textit{$^{a\ddag}$}}\vspace{0.5cm}

\noindent\textit{\small{\textbf{Received Xth XXXXXXXXXX 20XX, Accepted Xth XXXXXXXXX 20XX\newline
First published on the web Xth XXXXXXXXXX 200X}}}

\noindent \textbf{\small{DOI: 10.1039/c000000x}}
\vspace{0.6cm}

\noindent \normalsize{We study a mesoscopic model for the flow
of amorphous solids. The model is based on the key features identified at the microscopic level, namely periods of elastic deformation interspersed
with localised rearrangements of particles that induce long-range
elastic deformation. These long-range deformations are derived following
a continuum mechanics approach, in the presence of solid boundaries,
and are included in full in the model. Indeed, they mediate spatial
cooperativity in the flow, whereby a localised rearrangement may lead
a distant region to yield. In particular, we simulate a channel flow
and find manifestations of spatial cooperativity that are consistent
with published experimental observations for concentrated emulsions
in microchannels. Two categories of effects are distinguished. On
the one hand, the coupling of regions subject to  different shear rates, for instance,leads to finite shear rate fluctuations in the seemingly
unsheared {}``plug'' in the centre of the channel. On the other
hand, there is convincing experimental evidence of a specific rheology
near rough walls. We discuss diverse possible physical origins for
this effect, and we suggest that it may be associated with the bumps
of particles into surface asperities as they slide along the wall.}
\vspace{0.5cm}
\footnotetext{\textit{$^{a}$~Laboratoire Interdisciplinaire de Physique, Universit� Joseph Fourier
Grenoble, CNRS UMR 5588, BP 87, 38402 Saint-Martin d'H�res, Franc}}

\tableofcontents{}

\section{Introduction}

The flow of simple fluids can be described microscopically as a succession
of local, independent processes: collisions in the kinetic theory
picture or hopping events in the classical Eyring description. As
the temperature is lowered, or as the density increases, these processes
tend to become more collective, with a dynamical length scale that
increases as the glass transition is approached \cite{Berthier2005a,heussinger2010fluctuations}.
Eventually, the liquid falls out of equilibrium and acquires a nonzero
shear modulus on any finite time scale, as well as a yield stress
that must be overcome in order to initiate the flow. Similar changes
take place in athermal materials when the jamming point is crossed
following an increase of density. It is now quite well established
\cite{Rodney2011Review} that the flow mechanisms of such
amorphous solids are different in essence from those of liquids, as
they involve elastic interactions (shear waves) that are transmitted
through solids, but not through fluids. This results in nonlocal effects
in the flow of soft jammed/glassy materials, contrasting the case
of a simple fluid.

In fact, the flow of these materials bears notable similarities with
the dynamics of earthquakes \cite{Dahmen2011}, in that it features
a solid-like behavior at rest and local yielding above a given applied
stress. Yielding is characterized by the emergence of local 'shear
transformations' involving a few particles \cite{Argon1979}, associated
with a local fluidisation of the material. These structural rearrangements,
hereafter named plastic events, and also often referred to as shear transformations, or shear
transformation zones, in the literature,   induce long-range deformations. The microscopic details vary to some
extent with the particular nature of the material. In the case of
foams, they are identified as T1 events, in which the local change
of first neighbors is mediated by an unstable stage with four bubbles
sharing one vertex. In colloidal pastes and in atomic systems, they
involve relative displacements of limited magnitude within a small
group of atoms, which lead to a new equilibrium configuration that
is related to the original one by a shear deformation. In all cases
the stress that was originally supported by the particles partaking
in the plastic event is transmitted to the surrounding medium, which
behaves as an elastic continuum. The robustness of the above scenario
for an extremely wide range of materials is striking. Ample evidence
of the local plastic events and their long-range effects is indeed
provided both by experiments using diverse materials and simulations
\cite{Lemaitre2007,Tsamados2008,Falk1998}.

In the last two decades the modelling of flow in amorphous systems
has evolved along two distinct, but related, lines. First, several
models have been proposed that incorporate the flow scenario in an
average description. These models, among which the shear transformation
zone \cite{stz-review} and the soft glassy rheology (SGR) \cite{Sollich1997,Sollich1998}
models are the most sophisticated examples. 
Other simplified models falling into the same category are the fluidity
model \cite{Hebraud1998} or the very simple $\lambda$ model \cite{Coussot2002} describe the average
evolution of a population of flow defects under an imposed strain
rate in a mean-field-like manner. The effect of elastic interactions
between these defects is not directly accounted for, but enters the
models indirectly \emph{via} the introduction of parameters such as
an effective temperature associated with the mechanical noise. These
approaches have been remarkably successful in describing at least
some aspects of steady state flow curves, e.g., the existence of a
yield stress and the low shear rate behaviour, as well as transient
or oscillatory response in various systems, from metallic glasses
to foams or colloidal pastes. However, due to their intrinsic mean
field nature, fluctuations and spatial correlations in the flow are
discarded. Also, in their most simplified version they are unable
to account for heterogeneities and strain localisation. To capture
the latter phenomenon, extensions of the models have introduced a
coupling between the mean-field description and a diffusive behaviour
of the effective temperature, which again can be understood as a consequence
of the non-local interaction between elementary flow events \cite{Manning2007,Fielding2009,Bocquet2009}.

An alternative line of modelling consists in implementing numerically
the scenario of plastic events interacting through an elastic continuum
in the form of a discrete lattice model. Such an approach was pioneered
by Chen, Bak and Obukhov, in a model 
 initially proposed for the description of
earthquakes
\cite{Chen1991}, and by Argon and Bulatov \cite{Bulatov1994,Bulatov1994a,Bulatov1994b}.
A number of similar \emph{mesoscopic} models based on the same physical
scenario, but with different implementations, have been proposed and
studied in the literature \cite{Baret2002,Picard2005,Homer2009}.
The models are able to produce flow curves sharing similarities with
those observed experimentally, although significant differences are
revealed by closer inspection; they can account for strain localisation
and its dependence on the local dynamical rules \cite{Martens2012,Homer2009},
and allow one to explore the influence of parameters such as ageing
or temperature. They also reproduce the dynamical heterogeneities
observed in the flow, and their variation with strain rate\cite{Martens2011}.
However, these comparisons have generally remained qualitative, since
the models are in general rather schematic, ignoring in particular
tensorial aspects or convection.

In this contribution, we present a detailed study of a mesoscopic
model that incorporates these elements in a manner that allows a comparison
with experimental data obtained in simple geometries. In particular,
we will focus on the channel flow geometry and show that the model
captures experimental observations, including the fluctuations in
the local shear rates arising even in seemingly quiescent regions.
Such fluctuations are the hallmark of non-locality and spatial cooperativity
in the flow, which can give rise to spectacular long-range fluidisation
phenomena.

Section \ref{sec:Model_description} introduces the continuum mechanics-based
description of a plastic event and presents our mesoscopic model.
Details of its numerical implementation are also provided. In Section
\ref{sec:Model_fitting}, we fit the parameters of the model to experimental
data for concentrated emulsions taken from the literature,
 and we present the general features observed in our numerical simulations
of a channel flow. The last two sections focus on the manifestations
of spatial cooperativity in this particular geometry: Section \ref{sec:Bulk_cooperativity}
tackles cooperativity in the bulk, whereas some aspects of the specific
rheology near a wall are addressed in Section \ref{sec:Wall_rheology}.
A shorter account of some of these results has been described in Ref.\cite{Nicolas2013}.

\section{Continuum-mechanics based description of plastic events and presentation
of the mesoscopic model\label{sec:Model_description}}

Under homogeneous driving conditions, simple fluids flow homogeneously.
Amorphous solids, on the other hand, exhibit localised plastic events
when they are forced to flow \cite{Argon1979,PRINCEN1985,Maloney2006,Lemaitre2007,Amon2012},
associated with local shear transformations. In this section, we use
an approach rooted in continuum mechanics to describe the
effect of a plastic event on the surrounding (elastic) medium, along
with its time evolution. Then, we show how these results are integrated
into a mesoscopic model. The presentation of the model is brought
to completion by the choice of relevant probabilities for the onset
and end of a plastic event.  This section extends and details previous presentations in Ref. \cite{Picard2005,Nicolas2013}.

\subsection{Description of a plastic event}

Consider a rectangular system described by Cartesian coordinates $\left(x,y\right)$,
where $x\in\left[0,L_{x}\right]$ and $y\in\left[0,L_{y}\right]$
are the streamwise and crosswise coordinates, respectively. Should
the system be unbounded, the following results will be applicable,
provided that one takes their $L_{x}\rightarrow\infty$ and $L_{y}\rightarrow\infty$
limits. Otherwise, periodic boundary conditions are assumed, for the
time being.

On account of the solidity of the material (which is preserved at low shear rate), the response of the system to a perturbation
can be modelled by Hooke's law, whereby the local elastic stress $\boldsymbol{\sigma}^{el}$
is related to the local (deviatoric) strain $\boldsymbol{\epsilon}$\emph{
via} $\boldsymbol{\sigma}^{el}=\mathbf{C}\boldsymbol{\epsilon}$,
where $\mathbf{C}$ is the stiffness matrix. Before a perturbation
(here, a plastic event) sets in, mechanical equilibrium requires that:
\begin{equation}
\nabla\cdot\left(\mathbf{C}\boldsymbol{\epsilon}^{\left(0\right)}\right)-\nabla p^{\left(0\right)}=0,\label{eq:Mech_eq_0}
\end{equation}
where $p$ is the pressure, and the $\left(0\right)$ superscripts
denote the initial state. In the following, the material will be considered
incompressible, which implies that the displacement field $u$
obeys $\nabla \cdot u=0$, and isotropic, so that the
elastic stress can be written, in condensed notations, as 
\begin{equation}
\boldsymbol{\sigma}^{el}=\mu\left(\begin{array}{c}
\epsilon_{xx}-\epsilon_{yy}\\
\epsilon_{yy}-\epsilon_{xx}\\
2\epsilon_{xy}
\end{array}\right)=2\mu\left(\begin{array}{c}
\epsilon_{xx}\\
-\epsilon_{xx}\\
\epsilon_{xy}
\end{array}\right)\equiv2\mu\boldsymbol{\epsilon},\label{eq:stress_eq_strain}
\end{equation}
where $\mu$ is the shear modulus.

Clearly, Hooke's law will only hold within a certain limit. Indeed,
when the configuration is too strained locally, say, in a region $\mathcal{S}^{\left(0\right)}$,
particles rearrange so that the system is brought to a new local minimum:
this is a plastic event. While this rearrangement occurs, the memory
of the reference elastic configuration is lost, and, consequently,
the local elastic stress vanishes. The region undergoing the rearrangement
is therefore liquid-like and its stress will be mainly of dissipative
origin. Following this line of thinking and neglecting inertia, the
force equilibrium during the plastic rearrangement reads 
\begin{alignat}{1}
\begin{cases}
\nabla\cdot\boldsymbol{\sigma}^{diss}-\nabla p=0 & \text{ in region }\mathcal{S},\\
2\mu\nabla\cdot\boldsymbol{\epsilon}-\nabla p=0 & \text{ outside region }\mathcal{S}.
\end{cases}\label{eq:ContMech_globalEq}
\end{alignat}
 Notice that the boundaries of the plastic region shall be deformed
during the event and $\mathcal{S}$ refers to the \emph{deformed }region.
In Eqs.\ref{eq:ContMech_globalEq}, the dissipative stress $\boldsymbol{\boldsymbol{\sigma}^{diss}}$
was supposed to be mainly concentrated in the rearranging region.
For simplicity, we further assume that dissipation is linear with
respect to the strain rate, \emph{viz.} $\boldsymbol{\sigma}^{diss}=2\eta_{eff}\boldsymbol{\dot{\epsilon}}$.
This linearity is naturally to be understood as a simplification,
and not as a claim of the existence of some universality regarding
the dissipative mechanism (see Ref.\cite{LeMerrer2012} for a non-linear
law in the case of a foam). In addition to Eqs.\ref{eq:ContMech_globalEq},
force balance requires the continuity of the stress all along
the boundary of region $\mathcal{S}$. If $\mathcal{S}$ is small
enough so that the (plastic) deformation rate in this region can be
considered homogeneous, \emph{viz.}, $\boldsymbol{\dot{\epsilon}}\left(r\right)\equiv\boldsymbol{\dot{\epsilon}}^{pl}$
for $r\in \mathcal{S}$, the continuity of the stress
all along the boundary $\partial\mathcal{S}$ of the plastic inclusion leads
to: 
\begin{eqnarray}
\boldsymbol{\dot{\epsilon}}^{pl} & = & \frac{1}{\tau}\boldsymbol{\epsilon}_{\partial\mathcal{S}},\label{eq:eps_dot}
\end{eqnarray}
 where $\boldsymbol{\epsilon}_{\partial\mathcal{S}}$ refers to the
(elastic) strain on the outer boundary $\partial\mathcal{S}$. The time scale $\tau\equiv\frac{\eta_{eff}}{\mu}$
for the viscous dissipation of the elastic energy has been made apparent. 

The \emph{leading-order} response of the system to the plastic event
immediately follows from Eq. \ref{eq:eps_dot}: it simply comes down
to a (plastic) strain rate $\boldsymbol{\dot{\epsilon}}^{pl}\left(t\right)=\frac{1}{\tau}\boldsymbol{\epsilon}_{\partial\mathcal{S}}^{\left(0\right)}$
affecting only region $\mathcal{S}$. In an unconstrained environment,
the inclusion would therefore undergo a deformation $\boldsymbol{\dot{\epsilon}}^{pl}dt$
in a time interval $dt$.

However, since the inclusion is embedded in a solid,
the latter reacts to this plastic strain: supplementary elastic
stress and pressure fields, $\boldsymbol{\dot{\sigma}}^{\left(1\right)}dt=2\mu\boldsymbol{\dot{\epsilon}}^{\left(1\right)}dt$
and $\dot{p}^{\left(1\right)}dt$ respectively, are thereby induced
in the medium %
\footnote{This deformation will, in turn, affect the plastic deformation rate
$\boldsymbol{\dot{\epsilon}}^{pl}$, but these higher order
effects are neglected here.%
}. The derivation of the fields $\boldsymbol{\dot{\sigma}}^{\left(1\right)}$
and $\dot{p}^{\left(1\right)}$ is presented in the next subsection.
For the time being, let us remark that, thanks to the linearity of the
equations, one can express the induced stress on the boundary $\partial\mathcal{S}$
as 
\begin{equation}
\boldsymbol{\dot{\sigma}}_{\partial\mathcal{S}}^{\left(1\right)}=2\mu\boldsymbol{\mathcal{G}}_{0}\boldsymbol{\dot{\epsilon}}^{pl},\label{eq:sigma_ds1}
\end{equation}
where $\boldsymbol{\mathcal{G}}_{0}$ is a yet unknown tensor. Now,
since the response of the solid is a \emph{reaction }to an imposed
shear strain $\boldsymbol{\dot{\epsilon}}^{pl}dt$, it will \emph{oppose}
it, at least in the direct vicinity of the inclusion. Therefore, one
expects the eigenvalues of $\boldsymbol{\mathcal{G}}_{0}$ to
be \emph{negative}. Inserting Eqs. \ref{eq:stress_eq_strain} and
\ref{eq:eps_dot} into Eq. \ref{eq:sigma_ds1} yields, after simplification:
\begin{equation}
\boldsymbol{\dot{\epsilon}}_{\partial\mathcal{S}}^{\left(1\right)}\left(t\right)=\frac{\boldsymbol{\mathcal{G}}_{0}}{\tau}\boldsymbol{\epsilon}_{\partial\mathcal{S}}\left(t\right)\label{eq:eps_dot_bis}
\end{equation}
Equation \ref{eq:eps_dot_bis} expresses the fact that, up to a (potentially
time-dependent) shape prefactor $\boldsymbol{\mathcal{G}}_{0}$, the
force driving the rearrangement is the elastic stress imposed on $\mathcal{S}$
by the rest of the system, and that, in opposing this force, dissipation
sets a \emph{finite} timescale $\tau$ to this plastic transformation%
\footnote{The \emph{finite }duration of a plastic rearrangement, which is neglected
in the Soft Glassy Rheology model \cite{Sollich1997}, the Kinetic
Elastoplastic model \cite{Bocquet2009}, as well as in the mesoscopic
models of Refs.\cite{Baret2002,Homer2009}, might play a crucial role
in the compressed exponential relaxation of different soft materials.
For details, see Ref.\cite{Bouchaud2001}.%
}.  Cloitre and co-workers\cite{Cloitre2003} suggested that the duration of
a rearrangement in soft colloidal pastes coincides with the shortest
structural relaxation time $\tau_{\beta}$, which also results from
a  ``competition between elastic restoring forces and interparticle
friction'' , and experimentally confirmed the
proposed scaling $\frac{\eta_{eff}}{\mu}$ for the latter time (where $\eta_{eff}$ is determined by the dissipation within lubrication films). This scaling was also used to collapse flow curves onto a single master
curve, which bolsters its relevance for the rheology of these materials.

\subsection{Calculation of the elastic deformation induced by a single plastic
event (2D, tensorial)}

Reference \cite{Picard2004} proposed a method to derive the fields
$\boldsymbol{\dot{\epsilon}}^{\left(1\right)}dt$ and $\dot{p}^{\left(1\right)}dt$
induced by the plastic strain $\boldsymbol{\dot{\epsilon}}^{pl}dt$,
in a simplified context. First, one considers the limit of an
infinitely small plastic inclusion $\mathcal{S}$, $\boldsymbol{\epsilon}^{pl}\left(r\right)\rightarrow\boldsymbol{\epsilon}^{pl}a^{2}\delta\left(r-r_{0}\right)$,
where $r_{0}$ is the centre of region $\mathcal{S}$, and $a$, the typical linear size of $\mathcal{S}$. (The dots indicating derivation
w.r.t. time are omitted in this section). Secondly, by virtue of the linearity of the equations,
the inclusion applies a stress $\boldsymbol{\sigma}^{inc}=2\mu\alpha\boldsymbol{\epsilon}^{pl}$
on its surrounding, where $\alpha$ is a scalar (instead of a tensor) because of symmetry
arguments. At the expense of a renormalisation of the timescale $\tau$ appearing in the definition of
$\boldsymbol{\dot{\epsilon}}^{pl}$, Eq. \ref{eq:eps_dot}, \emph{viz.} $\tau\equiv \alpha^{-1}\frac{\eta_{eff}}{\mu}$, 
we can consider that $\alpha=1$. Mechanical equilibrium in the solid then
reads:
\begin{eqnarray}
2\mu\nabla\cdot\left[\boldsymbol{\epsilon}^{\left(1\right)}\right]-\nabla p^{\left(1\right)} & = & 2 \mu \nabla\cdot\left[\boldsymbol{\epsilon}^{pl}a^{2}\delta\left(r-r_{0}\right)\right]\label{eq:Guillemette}
\end{eqnarray}
To pursue, Eq. \ref{eq:Guillemette} is solved with the help of the
Oseen-Burgers tensor $\boldsymbol{\mathcal{O}}$, expressed in Fourier
coordinates $\underline{q}\equiv\left(p_{m},q_{n}\right)$, where
$p_{m}\equiv\frac{2\pi m}{L}$ and $q_{n}\equiv\frac{2\pi n}{L}$,
with $m,n\in\mathbb{Z}$:{ 
\begin{equation}
\hat{\mathcal{O}}(\underline{q})=\frac{1}{\mu\underline{q}^{2}}\left(1-\frac{1}{\underline{q}^{2}}\underline{q}\otimes\underline{q}\right).\label{eq:Oseen_Burgers_tensor}
\end{equation}
 }The Oseen-Burgers tensor is the elementary solution in terms of
displacement $u$ of the equations $\left\{ 2\mu\nabla\cdot\epsilon-\nabla p=\delta\left(r\right),\,\mathrm{div}\left(u\right)=0\right\} $,
where the linearised deformation tensor obeys $\epsilon=\frac{\nabla u+\left(\nabla u\right)^{T}}{2}$,
with the boundary conditions specified above. Therefore,
\begin{equation}
u^{(1)}\left(\underline{q}\right)=2\mu\boldsymbol{\hat{\mathcal{O}}}(\underline{q})\cdot\left(-i\underline{q}\cdot\hat{\boldsymbol{\epsilon}}^{pl}\right).
\end{equation}
 Recalling Hooke's law, $\hat{\boldsymbol{\sigma}}^{\left(1\right)}\left(\underline{q}\right)=2\mu\left[i\frac{\underline{q}\otimes u^{(1)}+\left(\underline{q}\otimes u^{(1)}\right)^{T}}{2}-\hat{\boldsymbol{\epsilon}}^{pl}\left(\underline{q}\right)\right]$,
one finally arrives at: 
\begin{equation}
\left(\begin{array}{c}
\hat{\epsilon}_{xx}^{(1)}\\
\hat{\epsilon}_{xy}^{(1)}
\end{array}\right)\left(\underline{q}\right)=\boldsymbol{\hat{\mathcal{G}}}^{\infty}\left(\underline{q}\right)\cdot\left(\begin{array}{c}
\hat{\epsilon}_{xx}^{pl)}\\
\hat{\epsilon}_{xy}^{pl}
\end{array}\right)\left(\underline{q}\right)\label{eq:defG_inf}
\end{equation}
 where the elastic propagator $\boldsymbol{\hat{\mathcal{G}}}^{\infty}$
obeys: 
\begin{equation}
\boldsymbol{\hat{\mathcal{G}}}^{\infty}\left(\underline{q}\right)\equiv\frac{1}{\underline{q}^{4}}\left[\begin{array}{cc}
-(p_{m}^{2}-q_{n}^{2})^{2} & \,\,\,-2p_{m}q_{n}(p_{m}^{2}-q_{n}^{2})\\
-2p_{m}q_{n}(p_{m}^{2}-q_{n}^{2}) & -4p_{m}^{2}q_{n}^{2}
\end{array}\right].\label{eq:formula_G_inf}
\end{equation}
 Equations \ref{eq:defG_inf} and \ref{eq:formula_G_inf} express the elastic deformation field induced by a pointlike plastic
event in a system with periodic boundary conditions. The corresponding
stress field is straightforwardly obtained by multiplication with
the shear modulus $2\mu$.

In real space, the propagator expressed in Eq.\ref{eq:formula_G_inf}
has a four-fold angular symmetry and decays as $r^{-d}$, where $d=2$
is the spatial dimension. These properties are consistent with observations
from atomistic simulations \cite{Leonforte2005,Maloney2006,Lemaitre2007}
as well as experiments \cite{Schall2007}.

Note that the present treatment does not describe the dilational effects\cite{Bokeloh2011}
possibly taking place during plastic events. These effects may naturally
add quantitative corrections to the picture drawn here, but, along
with the associated flow concentration coupling \cite{Besseling2010}
and free volume diffusion mechanisms (see Ref. \cite{Manning2007} and references therein), they are probably
not of paramount importance in the high density-low temperature situations
considered here \cite{Talamali2012}, where such effects are not always
present \cite{Goyon2008a,Chaudhuri2012}. 

\begin{figure}[H]

\begin{centering}
{\begin{centering}
\includegraphics[width=8cm]{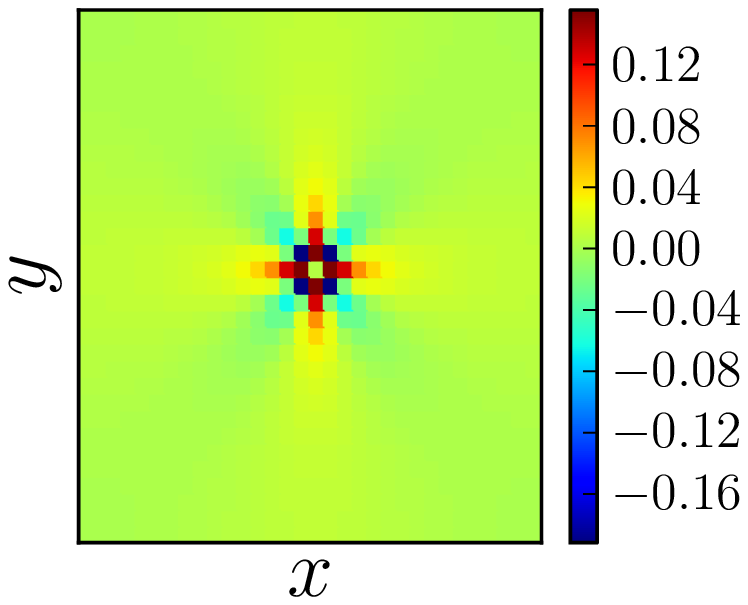}
\par\end{centering}

}
\par\end{centering}

\begin{centering}
{\begin{centering}
\includegraphics[width=8cm]{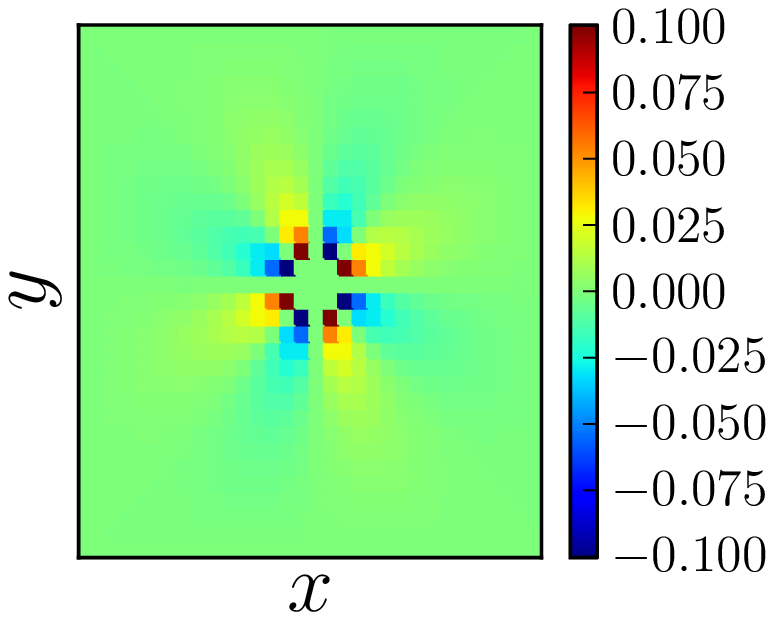}
\par\end{centering}

}
\par\end{centering}

\caption{Deformation field $\boldsymbol{\epsilon}^{\left(1\right)}$ induced
by a single plastic event $\epsilon_{xy}^{pl}$. Top: $\epsilon_{xy}^{\left(1\right)}$ component; bottom: $\epsilon_{xx}^{\left(1\right)}$ component. The values are normalised
by the absolute value of the locally induced deformation $\epsilon_{xy}^{\left(1\right)}$.
Because of the comparatively large (in magnitude) peak value at the
origin, the central block has been artificially coloured.\label{fig:Induced_def_field}}
\end{figure}

\subsection{Implementation of parallel confining walls}

In order to study a genuine channel geometry, the boundary conditions
need to be adapted to take into account two infinite parallel walls,
directed along $e_{x}$, bounding the flow, while keeping the periodicity
in direction $e_{x}$. The effect of the walls is modelled by imposing
a no-slip boundary conditions at their locations, in line with what
is commonly done in fluid mechanics.

\begin{figure}
\begin{centering}
\includegraphics[width=5cm]{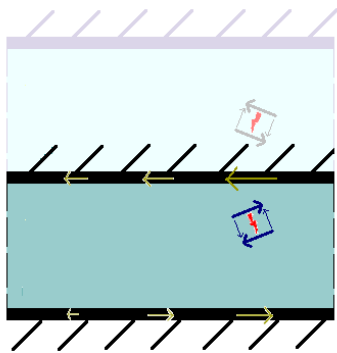}
\par\end{centering}

\caption{Sketch of the duplicated system.}
\end{figure}

To implement the no-slip boundary conditions, we extend the treatment
of Ref.\cite{Picard2004}: the system is duplicated in the direction
perpendicular to the walls, so that the region $y\in\left[0,L_{y}\right]$
describes the real system, while the region $y\in\left[-L_{y},0\right]$
is fictitious. For each plastic event (in the real system), a symmetric
'image plastic event' is created in the fictitious half. The $y$-component
of the velocity field is thereby cancelled at the walls. To remove
the \emph{x}-component of the velocity, adequate forces directed along
$e_{x}$ are added along the walls. These (fictitious) forces add
a corrective term $\hat{\epsilon}^{corr}$ to the deformation field
$\hat{\epsilon}^{\infty}$ obtained for periodic boundary:

\[
\hat{\epsilon}\left(\underline{q}\right)=\hat{\epsilon}^{\infty}\left(\underline{q}\right)+\hat{\epsilon}^{corr}\left(\underline{q}\right).
\]

The calculation of $\hat{\epsilon}^{corr}\left(\underline{q}\right)$
presented in Appendix \ref{sec:Appendix_Derivation_Correction_terms}
yields the following result:{ 
\begin{equation}
\left(\begin{array}{c}
\hat{\epsilon}_{xx}^{corr}\left(p_{m},q_{n}\right)\\
\hat{\epsilon}_{xy}^{corr}\left(p_{m},q_{n}\right)
\end{array}\right)=\left(\begin{array}{c}
\frac{-2p_{m}q_{n}^{2}}{\underline{q}^{4}}\left[i{\displaystyle \sum_{y}}\zeta_{\delta}\left(X\right)\,\mathcal{F}_{x}\epsilon_{xy}^{pl}(p_{m},y)+2{\displaystyle \sum_{y}}\xi_{\delta}(X)\mathcal{F}_{x}\epsilon_{xx}^{pl}(p_{m},y)\right]\\
\frac{q_{n}\left(p_{m}^{2}-q_{n}^{2}\right)}{\underline{q}^{4}}\left[i{\displaystyle \sum_{y}}\zeta_{\delta}\left(X\right)\,\mathcal{F}_{x}\epsilon_{xy}^{pl}(p_{m},y)+2{\displaystyle \sum_{y}}\xi_{\delta}(X)\mathcal{F}_{x}\epsilon_{xx}^{pl}(p_{m},y)\right]
\end{array}\right),\label{eq:eps_corr}
\end{equation}
}where $\sum_{y}$ denotes an integral over all streamlines $y=cst$
and $\mathcal{F}_{x}$ indicates a Fourier transformation along direction
\emph{x}. \emph{X }is used as a shorthand for {$\left(\frac{\pi y_{ev}}{L_{y}},\frac{p_{m}L_{y}}{\pi}\right)$}
and the analytical expressions of the functions $\zeta_{\delta}\left(X\right)$
and $\xi_{\delta}\left(X\right)$ can be found in Appendix \ref{sec:Appendix_Derivation_Correction_terms}.

Note that the corrective term couples the different Fourier modes
so that the translation invariance of the propagator $\mathcal{G}$
is broken (in the \emph{y}-direction). In particular, for a given
plastic strain, the local strain response now depends upon the distance
to the wall. The dependence on the distance for a plastic event $\left\{ \epsilon_{xx}^{pl}=0,\epsilon_{xy}^{pl}\neq0\right\} $
is presented in Fig.\ref{fig:Stress_release_vs_distance} for a system
that is coarse-grained into blocks of unit size (see next section).
In particular, one can see that the local strain relaxation induced
by a given plastic strain is around 35\% higher in the direct vicinity
of a wall than in the bulk case, owing to the vicinity of a solid
boundary.

\begin{figure}[H]

\begin{centering}
\includegraphics[width=8cm]{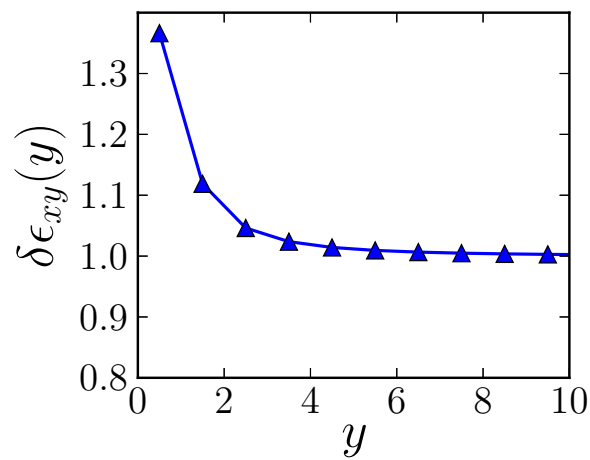} 
\par\end{centering}

\caption{\label{fig:Stress_release_vs_distance}Decrease $\left|\epsilon_{xy}^{\left(1\right)}\right|$
of the \emph{local} elastic strain induced by a given plastic
strain $\epsilon_{xy}^{pl}$ as a function of the distance \emph{y}
to the wall (expressed in block units, which is the only relevant
length scale). Values have been normalised to the 'bulk' value, that
is, the quantity measured infinitely far from the wall. }
\end{figure}

\begin{figure}[H]

\begin{centering}
\includegraphics[width=9cm]{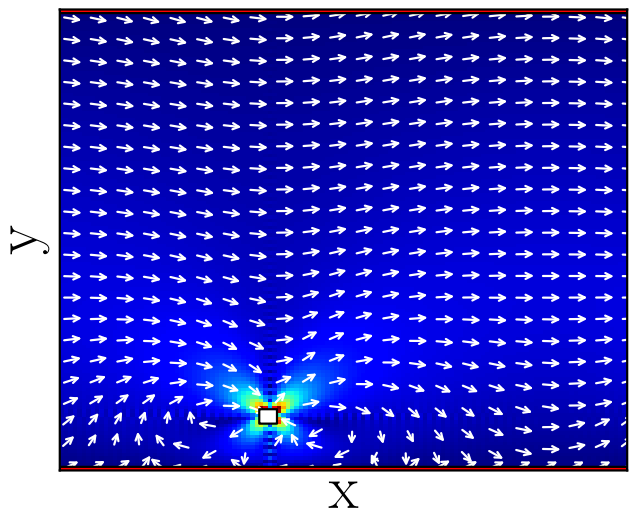} 
\par\end{centering}

\caption{\label{fig:Induced_Velocity_field}(Colour online) Displacement field
induced by a single plastic event (located in the white square). The
white arrows show the direction of the field, while the colour code
represents the displacement amplitude (brighter colours indicate higher
amplitude). Walls, drawn as red lines, are present at the top and
at the bottom of the system.}
\end{figure}

\subsection{Dynamics of the model and space discretisation}

At every point in space, the dynamics is governed by the following
equation, including both the external driving $\dot{\Sigma}^{ext}$
and the (local and nonlocal) stress redistribution due to plastic
events: 
\begin{equation}
\partial_{t}\boldsymbol{\sigma}\left(r\right)=\boldsymbol{\dot{\Sigma}^{ext}}\left(r\right)+\int\boldsymbol{\mathcal{G}}\left(r,r^{\prime}\right)\cdot2\mu\boldsymbol{\dot{\epsilon}^{pl}}\left(r^{\prime}\right)d^{2}r^{\prime},\label{eq:Master_eq}
\end{equation}
 where $\boldsymbol{\dot{\epsilon}^{pl}}\left(r\right)=\frac{\boldsymbol{\sigma}\left(r\right)}{2\mu\tau}$
if $r$ is in a plastic region, $\boldsymbol{\dot{\epsilon}^{pl}}\left(r\right)=0$
otherwise, and the propagator $\mathcal{G}$ takes into account both
the bulk (periodic) contribution and the corrections due to the presence
of walls. The plastic activity is determined by checking at every
time step, and at every point in space, the elements that undergo
a plastic event. The criterion for triggering plastic events will
be discussed in the next section. The time derivative in Eq. \ref{eq:Master_eq}
is handled numerically with a Eulerian procedure, with time step $dt\leqslant0.01\tau$.

The convolution part of Eq. \ref{eq:Master_eq} is most easily
solved in Fourier space, where the convolution turns into a product
involving the propagator derived previously (see Eqs. \ref{eq:formula_G_inf}
and \ref{eq:Master_eq}, for the two contributions to $\boldsymbol{\mathcal{G}}$).
To prepare the use a Fast Fourier Transform routine, the system is
spatially coarse-grained into a rectangular lattice of square-shaped
blocks of unit size. Physically, the size of the blocks should correspond
to the spatial extent of a plastic event.

Technically, the slow decay of $\hat{\sigma}_{xx}\left(q\right)$
with \emph{q} generates some irregularities in the computation of
the associated back-Fourier transform. Accordingly, for the sake of
precision, we use a finer mesh for the computation of the Fourier
transformations, i.e., we divide each block into four subblocks, so
that each plastic event now spans four subblocks. Thanks to this technical
trick, a smooth stress field is recovered, as shown in Fig.\ref{fig:Induced_def_field}.

Besides, mechanical equilibrium requires that the average of the shear
stress over any streamline (or any line with a given direction) be
homogeneous. However, the assumption of pointwise plastic events combined
with the discretisation of space is not entirely consistent, insofar
as it results in moderate violations of the aforementioned homogeneity,
when plastic events are far off the direction of macroscopic shear.
In order to restore homogeneity, an \emph{ad-hoc} shear stress is added globally to
every streamline. We have checked that this procedure has very little
effect on the results presented below.

\subsection{Coarse-grained convection}

Although the presence of a lattice precludes a rigorous implementation
of convection, a coarse-grained version can be introduced as follows:
The average velocity of each streamline in the flow direction is rigorously
calculated at each time step., viz.

\begin{eqnarray*}
\left\langle u_{x}\right\rangle _{x}\left(y_{0}\right) & \equiv & \frac{1}{L_{x}}{\displaystyle \int_{-\nicefrac{L_{x}}{2}}^{\nicefrac{L_{x}}{2}}}u_{x}\left(x,y_{0}\right)dx\\
 & = & \sum_{y_{ev}}\left[\mathrm{sign}\left(y_{0}-y_{ev}\right)\cdot\left(1-\frac{\left|y_{0}-y_{ev}\right|}{L_{y}}\right)+1-\frac{y_{ev}}{L_{y}}-\frac{y_{0}}{L_{y}}\right]\mathcal{F}_{x}\epsilon_{xy}^{pl}(m=0,y_{ev}).
\end{eqnarray*}
Details of the algebra are provided  in Appendix
\ref{sec:Derivation-of-the-velocity}.
The line displacement can thus be updated at each time step. Whenever
the displacement on a line grows larger than a multiple of the unit
block size, this line is incrementally shifted of the adequate number
of units, as a whole. In so doing, the regularity of the lattice is
preserved.

A technical detail might be worth mentioning: A na\"\i ve implementation
of the above method results in a violation of Galilean invariance,
insofar as lines with lower \emph{velocity} will be shifted less often
than others (artificial pinning) and therefore will tend to conserve
their neighbours (in the velocity gradient direction) for a longer
time - whereas the motion with respect to neighbouring lines should
be exclusively controlled by the local shear rate. It turns out that,
in a simple shear situation, the system is quite sensitive to such
a bias, which may lead 'pinned' lines to concentrate more plastic
activity. The practical solution to this issue consists in adding
a random offset displacement to \emph{all }streamlines, so that no
artificial pinning can occur.

\subsection{Probabilities for the onset and end of a plastic event}

So far, we have quantitatively described the effect of a plastic event and
detailed its derivation from rather well established principles. In
order to complete the model, criteria must now be fixed with regard
to the onset and termination of a plastic event. Since the mesoscopic
model is oblivious to the microscopic arrangement of the particles
and their stability, the criteria will obviously be somewhat arbitrary.
In the present model, they will involve two rates, $l\left(\sigma\right)$
and $e\left(\sigma\right)$, which govern, respectively, the transition
from the elastic to the plastic regime and the recovery of elastic behavior after
initiation of the plastic event,
\[
\text{elastic regime}\overset{l\left(\sigma\right)}{\underset{e\left(\sigma\right)}{\rightleftharpoons}}\text{plastic event}.
\]
 The use of rates introduces a simple element of stochasticity in
the model, and indirectly accounts for the variability of local environements.

Let us consider a mesoscopic region susceptible of undergoing a plastic
rearrangement. In the elastic regime, its configuration minimises
the potential energy, under the stress/strain constraints imposed
at the boundaries by the rest of the material. The minimum is stable
as long as $E-E_{constraint}\leqslant E_{a}$, where we assume the
existence of an average energy barrier $E_{a}$. In an Eyring-like
type of approach, the constraint is expressed as: $E_{constraint}\propto\sigma$,
where $\sigma$ is the local stress applied by the outer region, and
we take an activation volume equal to unity. Consequently, the rate
of plastic activation depends exponentially on the local stress. In
the following, we will use the expression 
\[
l\left(\sigma\right)=\Theta\left(\sigma-\sigma_{\mu y}\right)\exp\left(\frac{\sigma-\sigma_{y}}{x_{loc}}\right)\tau^{-1}
\]
 Three parameters have been introduced in this expression: $\sigma_{y}$
is the yield stress associated to the average energy barrier; $x_{loc}$
is a material-dependent activation temperature. Unlike the effective
temperatures used in the Soft Glassy Rheology model \cite{Sollich1998}
or the stress fluctuation approach \`a la Eyring\cite{Pouliquen2001},
$x_{loc}$ only accounts for local microscopic effects and does \emph{not}
include the local stress fluctuations due to stress redistribution,
i.e., mechanical noise: the latter should emerge naturally as a consequence
of long range interactions between events, within the framework described
above. Note that the limit $x_{loc}\rightarrow0$ of the activation
rate coincides with the usual von Mises yielding criterion, which 
states that the material yields if and only if $\sigma\geqslant\sigma_{y}$.
However, under shear stress, the effective lowering of energy barriers
results in the necessity to preserve the possibility of activated
events, even in materials usually referred to as athermal at rest.
For instance, the occurrence of rearrangements in granular matter
long after shear cessation \cite{Hartley2003} supports this claim,
although the physical reason for such rearrangements is far from clear.
Another possible justification for introducing $x_{loc}$ may be that
it effectively accounts for some dynamical disorder in the local yield
stress. Finally, we found that introducing this fluctuating, apparently
activated character in the yield criterion is the only way to obtain
flow curves in reasonable agreement with experimental data, as shown
below.

The parameter $\sigma_{\mu y}$ in the Heaviside function $\Theta$
is a critical stress, intended to be small $\sigma_{\mu y}\ll\sigma_{y}$
and below which no rearrangement can occur. Clearly, this is an \emph{ad
hoc} approach to conserve a finite macroscopic yield stress in the
limit of vanishing shear rate $\dot{\gamma}_{app}\rightarrow0$, when
no ageing process is explicitly taken into account. Note that Amon
\emph{et al.}\cite{Amon2012a}, in a paper investigating the behaviour
of granular matter on a tilted plane, recently called for a model
displaying two critical stresses, with a microfailure stress in addition
to the macroscopic one, although with a distinct definition.

A plastic event lasts until the local configuration gets trapped in
a new potential well. This trapping is expected to occur when the
local energy reaches low enough values, or, equivalently (recall that
$\sigma^{diss}\propto 2\mu \epsilon_{\partial \mathcal{S}}$, see Eq. \ref{eq:eps_dot}), when the dissipative stress
is dominated by the local elastic stress (which was neglected in Eq. \ref{eq:eps_dot}).
Consequently, we define an associated threshold for the recovery of
elastic stability, whose value is set to $\sigma_{\mu y}$ in order
to limit the number of parameters. Introducing a new intensive parameter
$x_{res}$, this allows us to write the rate at which elastic behaviour
is recovered as: $e\left(\sigma\right)=\exp\left(\frac{\sigma_{\mu y}-\sigma}{x_{res}}\right)\tau^{-1}$.

The definition of the rates $e$ and $l$ completes the description
of the model. At each time step, the probability of failure of an
elastic element is $l(\sigma)dt$, while the probability that a plastic
element becomes elastic again is $e(\sigma)dt.$

\section{Fitting of model parameters \& General observations in a channel
flow\label{sec:Model_fitting}}

In order to test the validity of the mesoscopic model presented in
the previous section, we start by fitting the model parameters by
comparing the flow curve obtained in simulations of a simple shear
setup to experimental results for a concentrated emulsion.

\subsection{Fitting of model parameters}

We use units of time and stress such that $\tau=1$ and $\sigma_{y}=1$,
and we set $\nicefrac{\mu}{\sigma_{y}}=1$ (note that the value of $\nicefrac{\mu}{\sigma_{y}}=1$ only contributes to rescaling 
the shear rate if convection is omitted).
The model then involves three parameters, $\sigma_{\mu y}$,
$x_{loc}$ and $x_{res}$.

In the following, our numerical simulations  are compared to experimental
data for concentrated oil-in-water emulsions collected by two different
groups, Goyon et al.\cite{Goyon2010} and Jop et al. \cite{Jop2012}.
The experimental systems are concentrated emulsions of $6-7{\mu m}$ droplets of silicon oil in a water-glycerol mixture at
an oil volume fraction $\phi \sim 0.75$ significantly larger than the jamming volume fraction.
 Both groups report a Herschel-Bulkley dependence of
the shear stress on the shear rate, that is, $\sigma=\sigma_{d}\left[1+\left(\tau_{HB}\dot{\gamma}\right)_{app}^{n}\right]$,
with an exponent $n\simeq0.5$ in both cases.

This Herschel-Bulkley law allows us to fit the remaining model parameters.
To do so, we simulate a simple shear flow by setting the driving force
to $\dot{\Sigma}_{app}=\mu\dot{\gamma}_{app}$ in Eq.\ref{eq:Master_eq},
with a stressless state as initial condition. By varying the parameters,
we find that the combination $\left\{ \sigma_{\mu y}=0.17,\, x_{l}=0.249,\, x_{e}=1.66\right\} $
provides a quite satisfactory fitting of the flow curve, as shown
in Fig.\ref{fig:Flow-curves}. Note that model units of time and stress
have been appropriately rescaled in the figure, to allow for comparison
with the experimental values. Of course, one may argue that the fitting
to the flow curve only loosely constrains the parameters, implying
that other combinations of parameters could yield the same result.
Nevertheless, we would like to mention that, when starting with a
moderately different set of parameters and fine-tuning it to better
match the data, we have recovered parameters similar to those selected.

\begin{figure}[H]

\begin{centering}
\includegraphics[width=8cm]{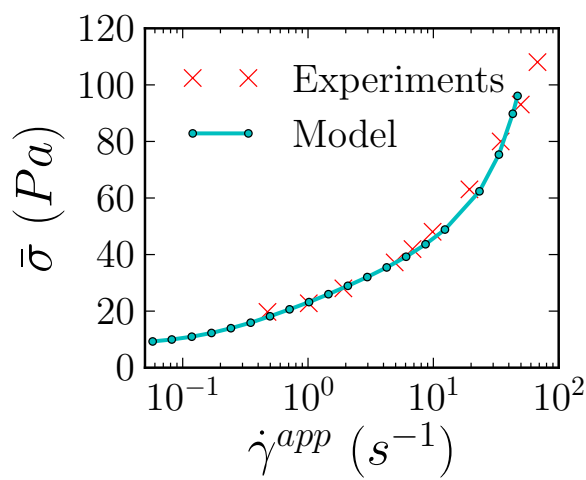} 
\par\end{centering}

\caption{(Crosses) Experimental and (dots) simulated flow curve. The experimental
were obtained by Goyon et al. for an emulsion of $\sim6.5\,\unit{\mu m}$ silicon oil droplets in a water-glycerin mixture at volume fraction $\phi=75\%$. The solid
line is a guide to the eye.\label{fig:Flow-curves} }
\end{figure}

\subsection{Channel flow: general observations}

Having set the model, we now turn to the specific case of channel
flow. Indeed, many intriguing experimental results have been reported
concerning the flow of soft jammed/glassy materials in that geometry
\cite{Goyon2008a,Goyon2010,Jop2012,Geraud2013,Isa2007,Isa2009}, which is also relevant for practical  applications, in particular in
the area of microfluidics.

First of all, it is important to realise that, unlike the 
simple shear case, the flow is pressure driven in a channel flow,
instead of being strain driven. Recalling that the driving $\Sigma^{ext}$
in Eq.\ref{eq:Master_eq} corresponds to the response of a purely
elastic solid, it immediately follows that: $\dot{\Sigma}^{ext}=0,\,\sigma_{xy}\left(x,y,t=0\right)=\nabla p\ \left(y-\nicefrac{L_{y}}{2}\right)$.
Note the streamline-averaged stress conserves a linear profile throughout
the simulation, because plastic events induce a homogeneous streamline-averaged
stress, owing to mechanical equilibrium.

We first discuss some general features of the flow of soft jammed solids
in that geometry.
Conspicuous is, in the first place, the presence of a  ''plug''
in the centre of the channel, i.e., a solid-like region in which the
material is convected, but not sheared. The plug can clearly be seen
in Fig.\ref{fig:Velocity_profiles}, which demonstrates a nice agreement
between the numerical and the experimental (time averaged) velocity
profiles across the channel. Note that showing the velocity differences
with respect to the maximal velocity across the channel obviates the
experimental issue of the determination of wall slip.

However, averaging over time masks the temporal fluctuations of the
flow. If one heeds the variations of the maximal streamline velocity of the simulated
flow with respect to time, flow intermittency becomes evident
\footnote{However, these fluctuations would presumably vanish in our model if the channel were of infinite length.}. 
This phenomenon is more acute for
narrow channels (see Fig.\ref{fig:Velocity_Oscillations}), in agreement
with results from numerical simulations regarding the effect of confinement
(see Ref. \cite{Chaudhuri2012} and references therein). Note that
flow intermittency, or {}``stick-slip'' behaviour, has often been
reported experimentally, but it has been interpreted in various ways
depending on the particular system under study: the creation and failure
of force chains is put forward in the case of granular matter \cite{Pouliquen1996,Gutfraind1996},
while variations in the local concentrations of colloids
and erosion by the solvent have been reported for concentrated colloidal
suspensions \cite{Isa2009}.

The spatial distribution of plastic events is also of interest. Indeed,
although the plug remains virtually still on average, sparse plastic
events are clearly seen to occur in that region, especially for narrow
channels, and, consequently, below the bulk yield stress. Therefore,
these plastic events essentially originate in cooperative effects,
via the redistribution of stress generated by distant plastic events.
Being of cooperative nature, the principal direction of their stresses
at the yielding point (the 'angle of yield' of the plastic event)
is broadly dispersed, since it is not strongly biased by a fixed
applied shear (see Fig.\ref{fig:YAP}).

\begin{figure}[H]

\begin{centering}
\includegraphics[width=8cm]{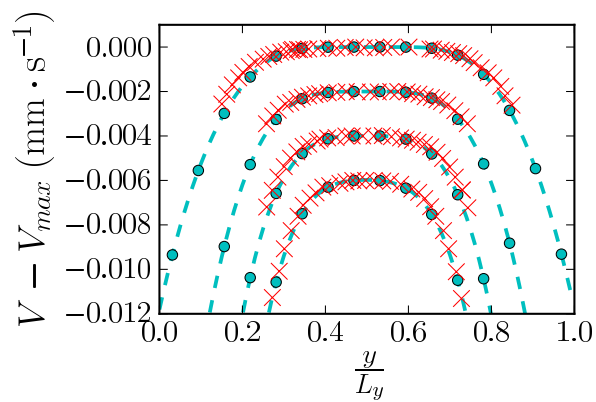} 
\par\end{centering}

\caption{(Crosses) Experimental and (dots) simulated velocity profiles, for stresses at the wall $\sigma_{w}=141\,\unit{Pa}$,
$188\,\unit{Pa}$, $235\,\unit{Pa}$, $282\,\unit{Pa}$, corresponding
to $\sigma_{w}$=0.36, 0.48, 0.60, 0.72 in model units, from top
to bottom. The
experimental data are a courtesy of Jop \emph{et al.}\cite{Jop2012}. The
model time and stress units have been rescaled to match the experimental
data. \label{fig:Velocity_profiles}}
\end{figure}

\begin{figure}[H]

\begin{centering}
\includegraphics[width=8cm]{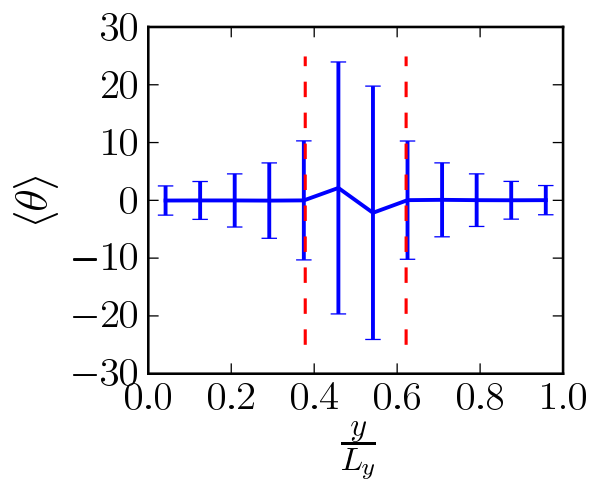} 
\par\end{centering}

\caption{
\label{fig:YAP}
Principal direction $\theta \in [-45^{\circ},45^{\circ}]$  of plastic events as a function of the position
in the channel. Channel width: 12. $\sigma_{w}$=0.6 in model units. The vertical dashed lines
delimit the 'plug', i.e., the region where $    |\langle \sigma_{xy} \rangle| \leqslant \sigma_{d}$.
The  bars  give  the standard deviation, 
$\pm \langle \langle \theta^{2}\rangle - \langle \theta \rangle ^{2}  \rangle$ 
}

\end{figure}

\begin{figure}[H]
\begin{centering}
\includegraphics[width=8cm]{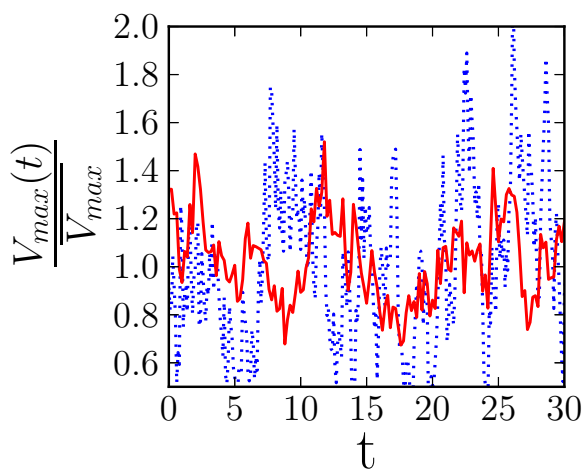} 
\par\end{centering}

\caption{Time variations of the maximal velocity in the channel, rescaled by
its average value over time, for two channel widths: (solid red line)
24 blocks, (dotted blue line) 6 blocks.\label{fig:Velocity_Oscillations}}
\end{figure}

\section{Cooperativity in the bulk flow: a manifestation of the coupling between
heterogeneous regions\label{sec:Bulk_cooperativity}}

\subsection{Origin and description of the nonlocality in the flow}

Spatial cooperativity is a hallmark of the flow of amorphous solids:
Because of the solidity of these materials, shear waves can propagate
in the bulk. Accordingly, a plastic event induces a long-range deformation
of the material and can thus set off other plastic events, possibly
triggering an avalanche. However, the channel geometry is particular
in that the driving is intrinsically inhomogeneous; therefore, cooperativity
couples regions (streamlines) subject to different stresses.

When considering a given region, one may then expect its behaviour
to differ from that it would have in a homogeneous flow. This
is a serious issue, since it undermines the paradigm that there exists
a constitutive equation relating the local shear rate to the local
shear stress, as explained by Goyon and colleagues \cite{Goyon2008a,Goyon2010}.
(Note, however, that doubts regarding the existence of a single flow
curve for concentrated emulsions had also been voiced earlier, following
experiments in a different geometry\cite{Salmon2003Emulsions}).

To rationalise the deviations that they observed experimentally, Goyon
et al.\cite{Goyon2008a} made use of a diffusion equation operating
on the local fluidity, that is to say, the inverse viscosity $ f\left(r\right)\equiv\nicefrac{\dot{\gamma}\left(r\right)}{\sigma_{xy}\left(r\right)}$:
\begin{equation}
\xi^{2}\Delta f-\left(f-f_{\mathrm{bulk}}\left[\sigma\left(y\right)\right]\right)=0,\label{eq:fluidity_diffusion}
\end{equation}
 where $f_{\mathrm{bulk}}\left(\sigma\right)$ denotes the fluidity
measured in a homogeneous flow at applied stress $\sigma$. The length
scale $\xi$ is a cooperative length, that scales with the particle
diameter \cite{Goyon2008a,Kamrin2012}. This diffusion equation is based
on the idea that plastically active regions will fluidise their neighbours,
and inversely. In Ref. \cite{Bocquet2009}, Bocquet and co-workers showed
that this equation can formally be derived from a Hebraud-Lequeux
fluidity  model \cite{Hebraud1998},  provided heterogeneities are taken into account. However,
the limitations imposed by analytical treatment required to cut off
the propagator beyond the first neighbours, and to consider the limit
of vanishing shear rate.

Nevertheless, Eq. \ref{eq:fluidity_diffusion} was  found to provide
a very satisfactory description of experimental and numerical data
in several cases\cite{Goyon2008a,Goyon2010,Jop2012,Chaudhuri2012,Geraud2013,Kamrin2012},
provided that the parameters, that is, the cooperativity length $\xi$
and the value $f_{\mathrm{wall}}$ of the fluidity at the wall, are
carefully fitted.

Assuming that this equation offers a valid first-order approximation of
the flow, we use it to assess the amplitude of the expected deviations from bulk behaviour.

To do so, we quantifying the extent of the coupling by estimating
the \emph{relative} deviations $\delta f\left(y\right)\equiv f\left(y\right)-f_{\mathrm{bulk}}$
of the fluidity. This defines a dimensionless number, the Babel number
$\mathrm{Ba}\equiv\frac{\delta f}{f}$. In Appendix \ref{sec:Appendix_Babel_number},
we show that, under the assumption of a Herschel-Bulkley constitutive
equation, Ba is of order $\left(\xi\frac{\Vert\nabla\sigma\Vert}{\sigma-\sigma_{d}}\right)^{2}$,
that is, $\left(\xi\frac{\Vert\nabla p\Vert}{\sigma-\sigma_{d}}\right)^{2}$
for a channel flow.

Noteworthy is the (quadratic) dependence of the Babel number on the stress
gradient, i.e., the pressure gradient in a Poiseuille flow. Indeed,
it is generally several orders of magnitude larger in microchannels
than in their larger counterparts, which explains why striking manifestations
of cooperativity have been observed only in the former. The Babel
number is also negligible in wide-gap Taylor-Couette geometry. For
instance, a rough estimation yields $\mathrm{Ba}\sim10^{-9}$ at most
in the wide-gap setup used by Ovarlez et al.\cite{Ovarlez2008}, where
no deviations from macroscopic rheology were reported. 

The denominator of Ba, $\left(\sigma-\sigma_{d}\right)^{2}$, also deserves
a comment: at high applied stresses, when the material is more fluid-like,
relative deviations become less significant. We should however say
that, to measure relative deviations, the absolute fluidity deviations
are divided by the fluidity, which gets large as $\sigma$ gets large.

\subsection{Nonlocal effects in the velocity profiles}

Following the above considerations, we expect deviations from macroscopic
rheology to increase with confinement, at fixed wall stress.

Indeed, Goyon's experiments on emulsions confined in microchannels with
\emph{smooth} walls tend to indicate that the deviations of the velocity
with respect to the bulk predictions follow such a trend. However,
overall, these deviations were found to be rather small. The mentioned
effect of confinement is also confirmed by Chaudhuri et al. with atomistic
simulations of a Poiseuille flow with biperiodic boundary conditions
with atomistic simulations \cite{Chaudhuri2012}. 

Figure \ref{fig:Velocity-profiles_deviations} shows a comparison
between the actual velocity profile obtained with simulations of the
mesoscopic model and the predictions from the (bulk) flow curve. As
in experiments, small deviations can be observed. For the extent of
these deviations to roughly match that in the experiments, the channel
width must be of order 7-10 block units. From this we deduce a first
estimate for the linear size $N_{\diameter}$ of a mesoscopic block
in terms of particle diameters: $N_{\diameter}\approx2$, which is comparable 
to experimental values found in the literature\cite{Schall2007}.

Let us now investigate how compatible our simulation results are with
the fluidity diffusion equation, Eq.\ref{eq:fluidity_diffusion}.
To solve Eq. \ref{eq:fluidity_diffusion}, two boundary conditions
are required: for symmetry reasons, we impose $f\left(y=0\right)=f\left(y=L_{y}\right)$,
and we set the fluidity at a point close to the wall to the value
measured in simulations. In addition, the shear-rate dependence of
the cooperativity length $\xi$ must be specified. Two possibilities
are considered in Fig.\ref{fig:Fluidity-profiles}: either, following
Goyon \emph{et al. \cite{Goyon2008a}, }$\xi$ is supposed independent
of the shear rate, i.e., $\xi=\xi_{0}$, or a power-law dependence
is assumed, $\xi\left(\dot{\gamma}\right)=\xi_{0}\left(\dot{\gamma}\tau\right)^{-\nicefrac{1}{4}}$,
where $\dot{\gamma}$ is the product of the local shear stress and
fluidity, as derived in Ref.\cite{Bocquet2009} in the limit $\dot{\gamma}\rightarrow0$,
and in reasonable agreement with the data of Ref.\cite{Jop2012} .
In both cases, $\xi_{0}$ is adjusted by a least square minimization.
Both cases give a reasonable fit, but neither matches our data accurately
over the whole range of applied pressures. We ascribe this defect,
among other details, to the approximation of long-range interactions
by a diffusive term, and to the neglect of fluidity fluctuations.

In Figure \ref{fig:RelDeviationVsBabel}, we assess the predictive
capability of the theoretically derived Babel number for our channel flow simulations by
plotting the $\frac{\delta f}{f}$ obtained in our simulations as
a function of $Ba=\left(\xi\frac{\Vert\nabla\sigma\Vert}{\sigma-\sigma_{d}}\right)^{2}$. It shows a global trend towards larger relative
deviations from macroscopic rheology for larger Ba, but the correlations
are poor. Nevertheless, one may expect Ba to still be a valid predictor
in practice, when widely different situations are considered.

\begin{figure}[H]

\begin{centering}
\includegraphics[width=8cm]{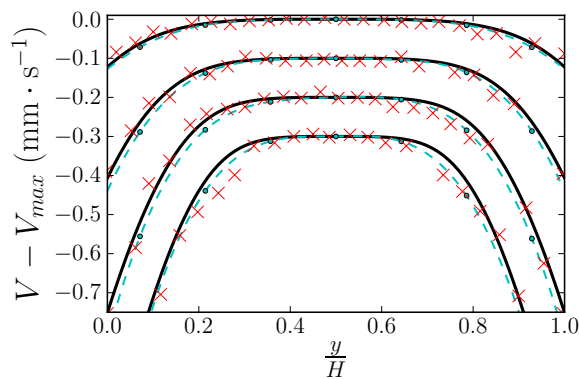} 
\par\end{centering}

\caption{\label{fig:Velocity-profiles_deviations}Velocity profiles across
the channel, for $\sigma_{w}=45,\,60,\,75,\,91\,\unit{Pa}$,
i.e., $\sigma_{w}=0.75,\:1.0,\,1.25,\,1.52$ in model units, from top to bottom: (dashed
line) simulation results, (solid line) predictions based on the numerical
bulk flow curve. The crosses are experimental data obtained by Goyon \emph{et al.} \cite{Goyon2010}. Note
that the curves have been shifted with respect to each other for clarity.}
\end{figure}

\begin{figure}[H]
\begin{centering}
\includegraphics[width=8cm]{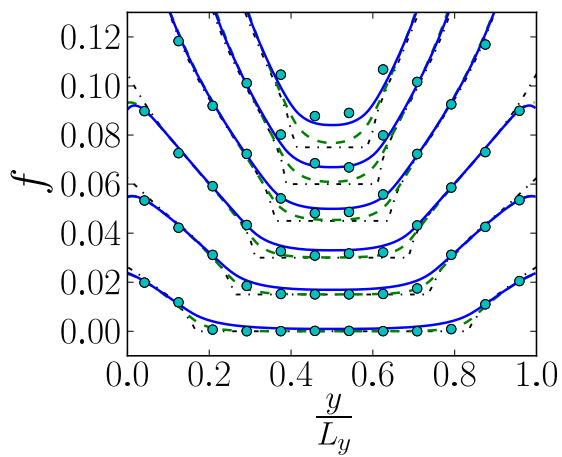} 
\par\end{centering}

\caption{\label{fig:Fluidity-profiles}Fluidity profiles for $N_{y}=12$, for $\sigma_{w}=0.20,\,0.28,\,0.36,\,0.48,\,0.60,\,0.72$ in model units. Filled
circles: numerical results, dashed green line: solution of Eq.\ref{eq:fluidity_diffusion}
with $\xi\left(\dot{\gamma}\right)=0.03702$, solid blue line: solution
of Eq.\ref{eq:fluidity_diffusion} with $\xi\left(\dot{\gamma}\right)=0.01146\,\dot{\gamma}^{-0.25}$.
The thin dash-dotted lines represent the bulk fluidity $f_{\mathrm{bulk}}$. Note that the curves are shifted with respect to each other for clarity.}
\end{figure}

\begin{figure}[H]
\centering{}\includegraphics[width=8cm]{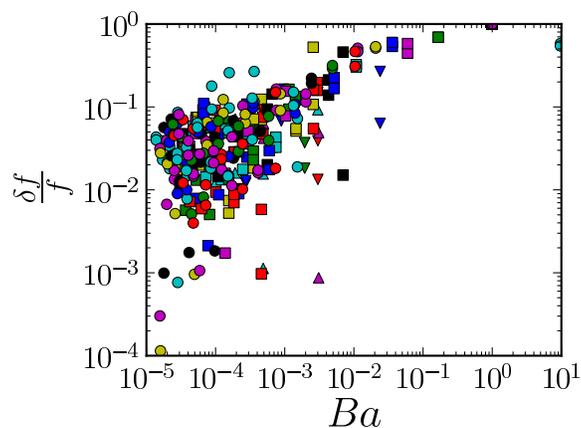} \caption{%
Relative deviations $\frac{\delta f}{f}$ of the local fluidity $f$
from the bulk fluidity $f_{bulk}\left(\sigma\right)$ measured in simulations, where $\sigma$
is the local shear stress, as a function of the estimated Babel number $Ba=\left(\xi_{0}\frac{\nabla\sigma}{\sigma-\sigma_{d}}\right)^{2}$.
We have set $\xi_{0}$ to 0.037 (see Fig.\ref{fig:Fluidity-profiles}).
Data only include regions where $\sigma>\sigma_{d}$, but cover various
applied pressures and channel widths: $\left(\blacktriangledown\right)$
6 blocks, $\left(\blacktriangle\right)$ 10 blocks, $\left(\blacksquare\right)$
16 blocks, $\left(\bullet\right)$ 24 blocks.\label{fig:RelDeviationVsBabel}
}
\end{figure}

\subsection{Shear rate fluctuations in the plug}

Quite recently, Jop and co-workers \cite{Jop2012} showed experimentally
that the seemingly quiescent plug in the centre of the microchannel
actually sustains finite shear rate fluctuations. This observation
is obviously consistent with the occurrence of sparse plastic events
in the plug, in our simulations.

To go further than this qualitative agreement, we directly compare
the local shear rate fluctuations $\delta\dot{\gamma}\left(x,y\right)=\sqrt{\left\langle \dot{\gamma}\left(x,y\right)^{2}\right\rangle -\left\langle \dot{\gamma}\left(x,y\right)\right\rangle ^{2}}$
to experimental data\footnote{Note that we have discarded the two curves corresponding to the lowest
applied pressures, which seem to plateau in the centre, because we
were not entirely sure of the accuracy of these measurements.%
}, with the parameters used to fit the associated velocity profiles
(see Fig.\ref{fig:Velocity-profiles_deviations}). Here, $\dot{\gamma}\left(x,y\right)$
is the local shear rate at point $\left(x,y\right)$; it is given
by $\dot{\gamma}\left(x,y\right)=2\left(\dot{\epsilon}_{xy}^{pl}\left(x,y\right)+\dot{\epsilon}_{xy}^{\left(1\right)}\left(x,y\right)\right)$
in the model and is therefore obtained directly, that is, without
deriving the velocity with respect to space. Figure \ref{fig:Shear-rate-fluctuations-vs-Exp}
presents the experimental shear rate fluctuation profiles and their
numerical counterparts for $N_{y}=16$ blocks crosswise. Semi-quantitative
agreement is observed in regions far from the walls - apart from the
large discrepancy at the highest applied pressure. The discrepancies
in the highly-sheared regions near the walls will be considered below.
It is interesting to note that the fitted channel size provides another
estimate for the size $N_{\diameter}$ of an elastoplastic block,
which agrees with the first one, $N_{\diameter}\approx2$. Figure
\ref{fig:Fluctuations_vs_channel_size} shows the dependence of the
shear rate fluctuations on the channel size for a given stress at
the wall. As expected from the expression of the Babel number, fluctuations
in the plug decay when the channel width is increased.

\begin{figure}[H]
\begin{centering}
\includegraphics[width=8cm]{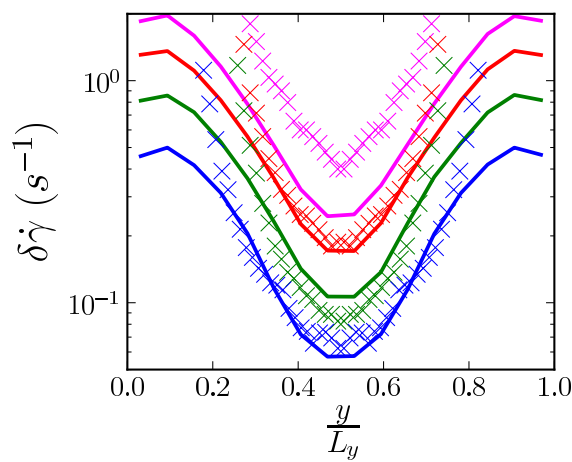} 
\par\end{centering}

\caption{\label{fig:Shear-rate-fluctuations-vs-Exp}Shear rate fluctuations
$\delta\dot{\gamma}\left(y\right)$ (averaged along the \emph{x}-direction),
for $\sigma_{w}=\, 141\,\unit{Pa},\ 188\,\unit{Pa},\ 235\,\unit{Pa},\ 
282\,\unit{Pa}$ (identical to Fig.\ref{fig:Velocity_profiles}),
from bottom to top. $\left({\color{red}\times}\right)$ Experimental
data collected by Jop \emph{et al.} \cite{Jop2012}, (\emph{solid
lines}) numerical results for $N_{y}=16$.}
\end{figure}

\begin{figure}[H]

\begin{centering}
\includegraphics[width=8cm]{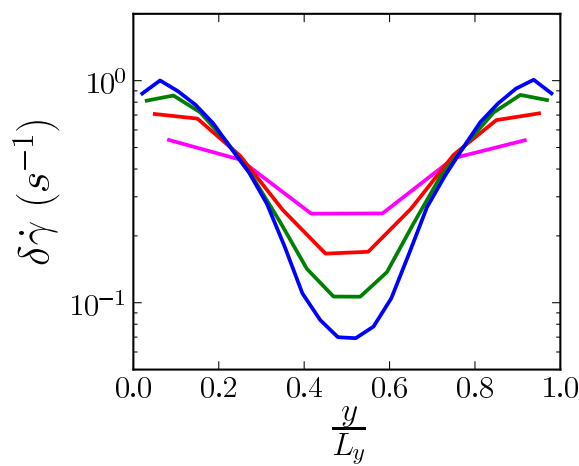} 
\par\end{centering}

\caption{\label{fig:Fluctuations_vs_channel_size}Shear rate fluctuation profiles
for a given stress at the wall, $\sigma_{w}=0.48$ in model units,
for different channel widths: (fuchsia) 6, (red) 10, (green) 16, and
(blue) 24 blocks, in descending order of minimal values.}
\end{figure}

Let us note that the data collected by Jop and co-workers suggested
a proportionality between the shear rate fluctuations and the local
fluidity, implying that both are indicators of the intensity of the
plastic activity. Figure \ref{fig:Plasticity_vs_Fluidity} shows that
the line-averaged plastic activity does indeed depend linearly on
the local fluidity in our channel flow simulations, despite some discrepancies
at low values of the fluidity, that is, probably in the plug. However,
the relation between the shear rate fluctuations and the mean fluidity
is much less clear (see Fig.\ref{fig:Fluctuations_vs_Fluidity}).

\begin{figure}[H]
\centering{}\includegraphics[width=8cm]{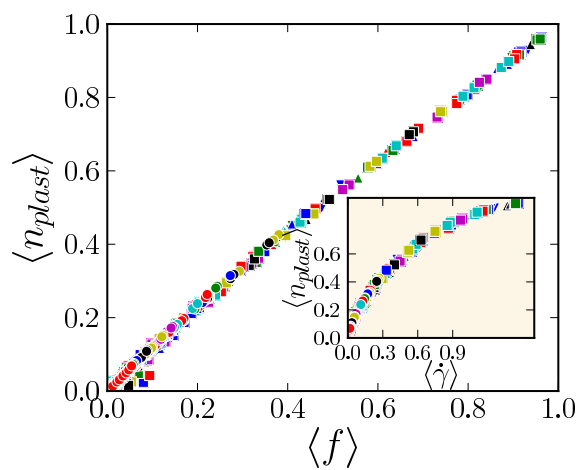} \textbf{\caption{%
Time-averaged fraction of plastic blocks $\left\langle n_{plast}\right\rangle $
on a given streamline as a function of the mean fluidity $\left\langle f\right\rangle $
on that line, for diverse applied pressures and various channel widths:
$\left(\blacktriangledown\right)$ 6 blocks, $\left(\blacktriangle\right)$
10 blocks, $\left(\blacksquare\right)$ 16 blocks, $\left(\bullet\right)$
24 blocks. \emph{Inset}: $\left\langle n_{plast}\right\rangle $ \emph{vs.}
the mean shear rate $\left\langle \dot{\gamma}\right\rangle $ on
the line. (Same symbols).\label{fig:Plasticity_vs_Fluidity}
}
}
\end{figure}

\begin{figure}[H]
\centering{}\includegraphics[width=8cm]{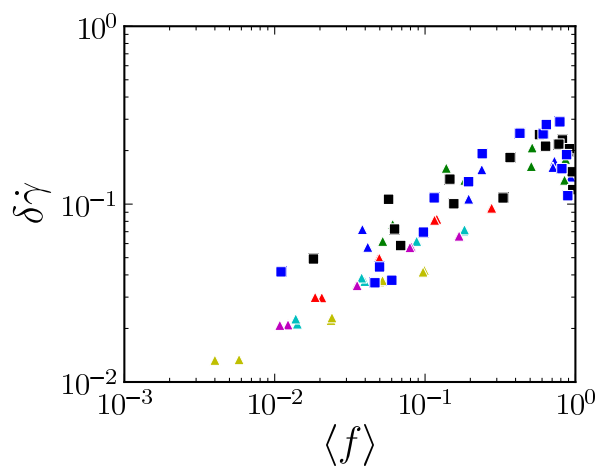} \textbf{\caption{%
Shear rate fluctuations $\delta\dot{\gamma}$ on a given streamline
as a function of the mean fluidity $\left\langle f\right\rangle $
on that line, for diverse applied pressures and various channel widths:
$\left(\blacktriangle\right)$ 8 blocks, $\left(\blacksquare\right)$
16 blocks, $\left(\bullet\right)$ 24 blocks.\label{fig:Fluctuations_vs_Fluidity}
}
}
\end{figure}

\section{A specific rheology near the walls?\label{sec:Wall_rheology}}

In the previous section, we have dealt with the flow cooperativity
associated with the coupling of heterogeneous streamlines, leaving
aside another potentially significant difference with bulk homogeneous
flow: the presence of walls bounding the flow, which is known to affect
the flow of diverse complex fluids: wormlike micellar solutions \cite{Masselon2010,Becu2004},
laponite \cite{Gibaud2008}, dense colloidal suspensions \cite{Isa2007},
etc. Indeed, Goyon et al. provided experimental evidence of the occurrence
of ample changes when rough walls are substituted for smooth walls
\cite{Goyon2008a}. Then, much larger deviations from bulk rheology
are observed, especially at high applied pressures, and these deviations
are maximal close to the walls, contrary to predictions based on the
Babel number.

\subsection{Weak deviations due to no slip boundary condition}

Remember that walls are described by a
no-slip boundary condition in our model. This condition results in a significantly
larger dissipation  during plastic events in their vicinity. Is
this sufficient to capture the very large deviations observed experimentally?

Figure \ref{fig:Local_flow_curve} shows the local flow curve for
the simulations. To decouple to a certain extent the problem of wall
rheology from the inhomogeneous driving, the latter being associated
with large values of $\mathrm{Ba}$, a relatively large channel is
considered here. For each value of the wall stress, the points with
the highest local shear rates in Fig.\ref{fig:Local_flow_curve} are
closest to the walls. We do observe some slight deviations%
\footnote{Nevertheless, replacing the no slip boundary condition with a periodic
boundary condition will play a role if the Babel number is large enough.
See Ref. \cite{Chaudhuri2012} for the effect of confinement on the
observed yield stress in a biperiodic Poiseuille flow. %
}, but they are clearly much weaker than in Goyon's observations (see
Fig.6 of Ref. \cite{Goyon2010} for instance). In this respect, they
much better describe the situation for smooth walls, which, at first,
might seem surprising given the no-slip boundary conditions. Yet,
in reality, the large slip observed at smooth walls is not expected
to give rise to significant changes: it only adds a simple global
translation to the complex velocity field obtained with no-slip boundary
conditions.

\begin{figure}[H]

\begin{centering}
\includegraphics[width=8cm]{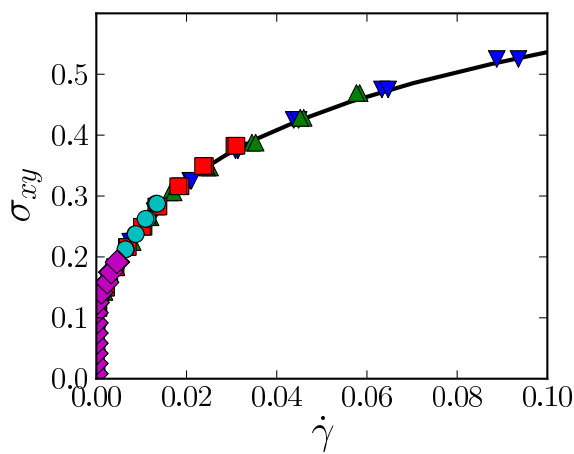} 
\par\end{centering}

\caption{Local shear stress as a function of the local shear rate, for various
applied pressures for a channel width of 24 blocks. The corresponding
stresses at the walls are: (\emph{purple rhombs}) $\sigma_{w}=0.2$,
(\emph{cyan dots}) $\sigma_{w}=0.2$, (\emph{red squares}) $\sigma_{w}\simeq0.4$,
(\emph{green upper triangles}) $\sigma_{w}\simeq0.5$, (\emph{blue
lower triangles}) $\sigma_{w}=0.6$.\label{fig:Local_flow_curve}}
\end{figure}

\subsection{Physical effect of rough walls}

As the deviations observed for rough walls are not
captured by a simple no slip boundary condition, we discuss here some 
physical mechanisms that may be responsible for the observed behaviour. 

First, the static structure near walls is known to differ from that
in the bulk. For smooth, or not too rough, boundaries, stratification
in layers is often reported over a distance of a few particle diameters
\cite{Ballesta2008,Mansard2012PhD}, though not systematically: Goyon
et al. \cite{Goyon2010} actually observed no such layering in their
experiments. Besides, the vicinity of a solid boundary hinders the
mobility of Brownian particles\cite{Pagac1996}. But these structural
changes for the material at rest imply a decrease of the fluidity
at the wall, as opposed to the enhancement that is experimentally
observed by Goyon \cite{Goyon2008a} and G\'{e}raud \cite{Geraud2013} at high enough stresses, i.e., where the largest deviations
occur. Alternatively, the specific behaviour at the wall is often rationalised
by the existence of a depleted 'lubrication layer' close to the wall,
as is often found in sheared dispersions \cite{yoshimura1988wall,Barnes1995,Franco1998,Meeker2004,Meeker2004long,Salmon2003,Becu2005}.
This phenomenon is more acute for deformable particles \cite{Franco1998} undergoing
high shear rates and/or high shear gradients; it generates an apparent
wall slip. However, at the very high concentrations investigated here,
owing to the large osmotic pressure, such a lubricating layer would have a
thickness of order $100\,\unit{nm}$ or less \cite{Salmon2003,Becu2005,Goyon2010}
(if the lubricating layer is composed of pure solvent). Effectively,
Goyon directly measured the concentration profile across the channel
and was not able to detect any significant variation. This finding
is corroborated by the absence of radial droplet migration for a similar
material in a Taylor-Couette cell, even at high shear rates, as reported
by Ovarlez et al. \cite{Ovarlez2008}. Adding that systems of soft
particles have a much weaker viscosity dependence on concentration
than their hard particle counterparts, effects of concentration variations
could be ruled out as regards Goyon's experiments. Nevertheless, we
attempted to simulate a less viscous layer close to the wall by decreasing
the yield stress of the associated mesoscopic blocks, but this only
had little effect on the rest of the system. Therefore, one is led
to seek another explanation. 

An aspect that has been overlooked so far is the reported observation
of wall slip in Goyon's, Geraud's and Jop and Mansard's works \cite{Goyon2008a,Goyon2010,Geraud2013,Mansard2012PhD},
both with smooth and rough walls. In order to extract information
that is relevant for the bulk flow, the authors measured the \emph{local
}velocities and shear rates in the channel by microscopic observation,
so that the occurrence of slippage should \emph{a priori} not affect
their results. Indeed, in presence of smooth surfaces, where wall
slip accounts for around 30\% of the maximal velocity at the typical
pressures applied by Goyon et al. \cite{Goyon2008a}, slip only results
in a global translation of the system, that leaves the local flow
curve strictly unaltered. For rough surfaces, let us first remark
that the presence of wall slip is more surprising, since roughened  surfaces
\footnote{Diverse methods are available for roughening a surface, such as sandblasting, covering it with sandpaper,
  or coating it with particles.}
 are often
used to strongly suppress, or eliminate, slip for the very same type
of materials, which is monitored by rheological measurements, and
then used as benchmarks for a system without slip \cite{Barnes1995,Mason1996JColl,Sanchez2001,Meeker2004long,Meeten2004}.
However, in several cases, measurements of local velocities in the
flow, either with microvelocimetry with fluorescent tracers \cite{Goyon2010,Goyon2008PhD,Geraud2013}
or through direct visualisation with confocal microscopy \cite{Jop2012,Mansard2012PhD},
demonstrate that concentrated emulsions may slip along rough surfaces
in microchannels. A seemingly quadratic \cite{Goyon2008PhD,Geraud2013},
or linear \cite{Mansard2012PhD}, dependence of the slip velocity
on the shear stress at the wall in reported in these cases. As a side note, let us
add that slip along a rough wall
is not restricted to the microchannel geometry: for instance, Divoux
\emph{et al.} showed with ultrasonic speckle velocimetry that another
yield stress fluid, namely carbopol, experiences a phase of total
slip in a Taylor-Couette rheometer whose cylinders had been coated
with sand paper \cite{Divoux2011Overshoot}. 

Now, when particles slide along a rough wall, they are expected to
bump into, and be deformed by, the surface asperities. In the case
of asperities that are large as compared
to the {}``particle'' size ($\sim60$ microns \emph{vs.} from a few to 20 microns),
this phenomenon is best exemplified by the spatiotemporal diagram
acquired with ultrasonic velocimetry for a carbopol microgel, Fig.6
of Ref.\cite{Divoux2011Herschel}, where one can see a large deformation
of the material that originates at the rotor and propagates almost
instantaneously into the bulk; this signal was interpreted by the
authors as the signature of a {}``bump'' into a surface protuberance.
Albeit less visible, this effect should also appear with walls characterised
by a smaller roughness, whereby rough walls in the presence of slip
act as sources of mechanical noise and cause deviations from bulk
rheology in their vicinity. This tentative scenario has the potential
to explain why deviations may, or may not, be observed in the vicinity
of rough surfaces: for instance, Goyon et al.\cite{Goyon2010} and
Ovarlez et al. \cite{Ovarlez2008}, as well as Seth and co-workers
\cite{Seth2012}, have reported that the local flow curves obtained
in wide gap Taylor-Couette or plate-plate geometries with rough walls
could be mapped onto the macroscopic flow curves; yet, they also indicated
that, in those cases, no evidence of wall slip could be found.  Very recently,
Mansard endeavoured to investigate the impact of wall
roughness by combining experiments and molecular dynamics simulations
\cite{Mansard2012PhD}. Nonmonotonic variations of the wall fluidity as a function of the
roughness were reported in the experiments, but the data did not allow for the extraction
of the parameters responsible for the deviations from
from macroscopic rheology. Nevertheless, he noted that {}`` the particles must
jump over the patterns {[}on the walls{]}. This effect induces the
rearrangements and increases the wall fluidity''. 

Naturally, this prompts the following question: what determines the occurrence of slip
along rough walls? This question lies far beyond the scope of the
present study. Let us simply note that in Refs. \cite{Goyon2008a,Goyon2010,Jop2012,Geraud2013}
the size of surface asperities was a couple of microns at most, that is,
significantly less than the typical {}``particle'' size, which  plausibly
favours slip, as well as the high shear rates experienced at the microchannel
walls. 
Nevertheless, recent theories of slip along smooth walls involved,
in addition, parameters such as the deformability of the droplets,
\cite{Meeker2004,Meeker2004long} and the particle-wall
interactions \cite{Seth2008}, not to mention the presumably significant impact of Brownian motion in cases where it is relevant \cite{Besseling2009,Ballesta2012}.
As far as we know, the somewhat daunting challenge to extend these
theories to the case of rough walls still awaits a successful accomplisher.

In the above discussion, we have carefully eluded the question of
the surface chemistry and its interactions with the particles. However,
Seth and co-workers showed that they can play a signifiant role; in
particular, for the yield stress fluid they studied, smooth attractive
surfaces were observed to induced deviations from macroscopic rheology
relatively far into the bulk, whereas smooth repulsive induced none
at all. 

Finally, we would like to mention another possible impact of the confinement
of the material between walls. The channel may be so narrow that the
layers where the specific wall rheology dominates start overlapping.
This situation, which is described as strong confinement,
is expected to occur when the channel width becomes of the order of,
or smaller than, the cooperativity length $\xi$. For the data of
Refs.\cite{Goyon2008a,Goyon2010,Jop2012} discussed above, this mechanism
is therefore not relevant.

\subsection{Fictitious plastic events along the wall as mechanical noise sources}

As we have already noted, nonlocal effects leading to deviations from
the macroscopic flow curve are often rationalised in terms of the
fluidity diffusion equation, Eq.\ref{eq:fluidity_diffusion} (see,
e.g., Ref.\cite{Goyon2008a,Goyon2010,Jop2012,Chaudhuri2012,Seth2012,Mansard2012PhD,Geraud2013,Kamrin2012}).
In this approach, the fluidity at the wall is needed as an input parameter, whose
precise value turns out to be determinant. Most likely, the suggested
mechanical noise at the walls would be hidden in that value. (Note
that, in Goyon \cite{Goyon2008PhD} the fluidity at rough walls, where
larger deviations are observed, is indeed larger than that for smooth
walls and larger than the bulk fluidity corresponding to the same
shear stress.)

Our mesoscopic model is also oblivious to the microscopic details
of the flow near a boundary and therefore cannot describe the effect
of wall slip along a rough wall without further input. Nevertheless,
since bumps act as sources of mechanical noise in the system, one
can attempt to account for their occurrence by adding fictitious plastic events along the walls. Note that this ad hoc treatment is
similar to imposing a wall fluidity larger than the bulk fluidity
as a boundary condition when solving the fluidity diffusion equation,
Eq. \ref{eq:fluidity_diffusion}.

More precisely, we modify the implementation of the model slightly,
so that a wall is now described as a line of plastically inert blocks:
the bottom wall will, for instance, occupy the portion of space $0\leqslant y\leqslant1$,
and the no slip boundary condition is imposed at its centre, i.e.,
$y=0.5$. On this line, a fraction of blocks is selected%
\footnote{Note that shuffling these blocks, i.e., selecting new random blocks
as noise sources, at low enough frequency hardly affects the results
presented below. %
} at random to act as sources of mechanical noise, that is, to mimic,
e.g., bumps of particles into surface asperities. To do so, they shall
release a constant plastic strain $\dot{\epsilon}_{fict}^{pl}$ per
unit time, along the direction of macroscopic shear (for simplicity).
We emphasise that mechanical equilibrium is  not violated
by the addition of these fictitious  plastic events.

Figure \ref{fig:Local-flow-curve_artificious} shows the local flow
curves obtained with this protocol. The observed deviations are qualitatively
similar to those reported by Goyon (see Fig.7 of Ref.\cite{Goyon2010}).
However, we must note that a rather intense mechanical noise is required
to get such deviations $\left(\dot{\epsilon}_{fict}^{pl}\approx5\right)$.
(As the value of $\dot{\epsilon}_{fict}^{pl}$ is arbitrary,
we do not   seek quantitative agreement with the experimental
data here). In addition, these fictitious  plastic events also alter the shear
rate fluctuation profile, as shown in Fig. \ref{fig:Shear-rate-fluctuation_artificious}.
Besides, a global increase of the fluctuations, the profile no longer
flattens in the vicinity of the walls, which renders it more consistent
with the experimental results of Jop and co-workers (collected in
a channel with rough walls).

\begin{figure}[H]

\begin{centering}
\includegraphics[width=8cm]{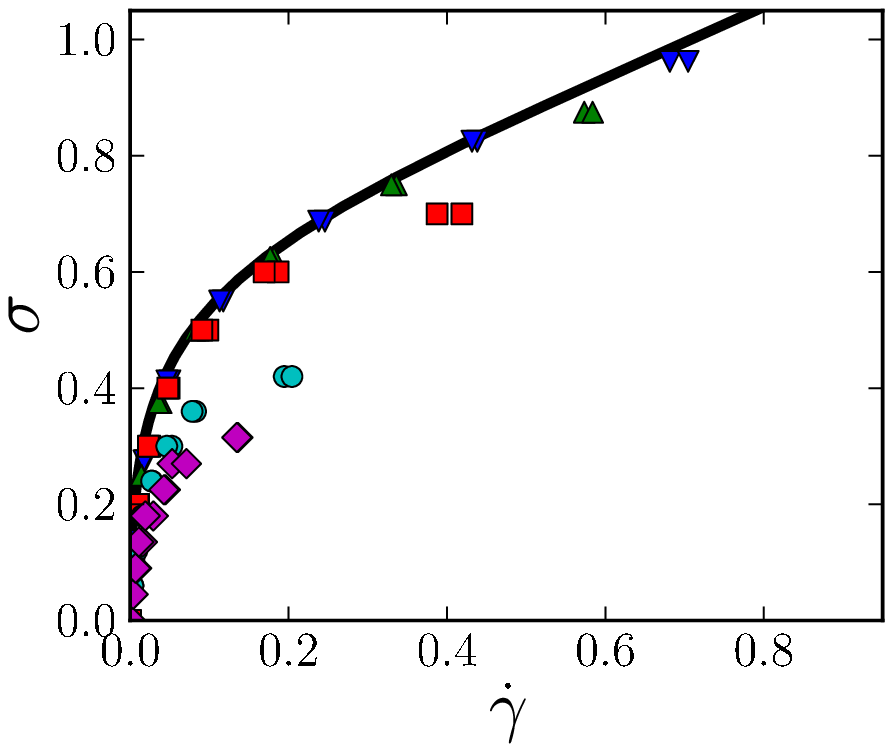} 
\par\end{centering}

\caption{\label{fig:Local-flow-curve_artificious}Local shear rate $\sigma\left(y\right)$
\emph{vs }local shear rate $\dot{\gamma}\left(y\right)$ (averaged
on streamlines $y=cst$) in the microchannel, when fictitious mechanical
noise sources of intensity $\dot{\epsilon}_{xy}^{fict\, pl}=\pm4.5$
are added on a fraction ($\nicefrac{1}{3}$) of blocks on the wall
lines. $\sigma_{w}$=$\left({\color{magenta}\blacklozenge}\right)$
$0.36$, $\left({\color{cyan}\bullet}\right)$ $0.48$, $\left({\color{red}\blacksquare}\right)$$0.8$,
$\left({\color{green}\blacktriangle}\right)$1.0, $\left({\color{blue}\blacktriangledown}\right)$$1.1$
in model units. Solid line: macroscopic flow curve. }
\end{figure}

\begin{figure}[H]

\begin{centering}
\includegraphics[width=8cm]{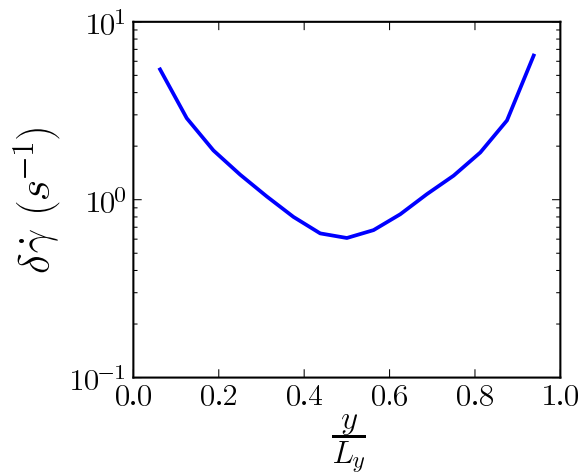} 
\par\end{centering}

\caption{\label{fig:Shear-rate-fluctuation_artificious}Shear rate fluctuation
profiles in the presence of fictitious plastic events along the walls.
A third of the blocks on wall lines have been randomly selected to
release a constant plastic stress $\dot{\sigma}_{xy}^{pl\, fict}=4.5$ per unit time. }
\end{figure}

\section{Conclusions \& Outlook}

In conclusion, we have derived analytical formulae from continuum
mechanics for the effect and time evolution of a plastic event occurring
in a two dimensional medium bounded by walls. We have integrated these
formulae in a lattice model for the flow of amorphous solids, in which
elastoplastic blocks receive stress from their surroundings and have
a certain probability to become plastic; the chosen form of probabilities
for the onset and end of a plastic event allowed us to match experimental
flow curves for concentrated emulsions. Then we turned to the simulation
of flow in microchannels, where the most prominent feature is the
existence of a seemingly unsheared {}``plug''. Remarkable manifestations
of spatial cooperativity in the flow had been unveiled experimentally,
and we proposed to distinguish those pertaining to cooperativity in
the bulk from those pointing to the specific rheology near a solid
boundary. For the former category, deviations of time averaged quantities
are generally weak, but could nevertheless be observed with our model.
More strikingly, shear rate fluctuations were observed in the plug,
consistently with experiments. As regards the specific wall rheology,
it turned out that imposing no-slip boundary conditions at the walls
in our model was not sufficient to capture the experimentally observed
phenomena. We discussed several possible physical origins for the
departure from the macroscopic behaviour observed, above all, in the
vicinity of rough surfaces; we insisted in particular on a tentative
scenario in which mechanical noise is created at the wall by, e.g.,
bumps of particles into surface asperities as they slide along the
wall. Finally, an ad hoc implementation of this mechanical noise was
attempted.

Concerning our mesoscopic model, several improvements can be considered.
First and foremost, regions undergoing plastic events are fluidised,
and the presence of fluid-like regions is expected to damp shear waves
and reduce cooperativity. This point is not taken into account in
the model. Also, the distinction between an activation temperature,
of noncooperative origin, and a more general effective temperature
will be worth further investigation, both for thermal and 'athermal'
soft solids under shear. In an unrelated way, it has been apparent
that, in spite of the vast amount of literature on the question of
slip for soft solids and the recent progress made in that respect\cite{Meeker2004},
the issue of   slip along a rough surfaces, and its consequences on the local fluidity, remains quite challenging.

\section*{Acknowledgments.} We thank T. Divoux, K. Martens, P. Chaudhuri,
S. Manneville, and M. Fardin for interesting discussions. JLB is supported by Institut Universitaire de France and
 by  grant ERC-2011-ADG20110209.

\bibliographystyle{rsc}
\bibliography{PhDbib}

\providecommand*{\mcitethebibliography}{\thebibliography}
\csname @ifundefined\endcsname{endmcitethebibliography}
{\let\endmcitethebibliography\endthebibliography}{}
\begin{mcitethebibliography}{77}
\providecommand*{\natexlab}[1]{#1}
\providecommand*{\mciteSetBstSublistMode}[1]{}
\providecommand*{\mciteSetBstMaxWidthForm}[2]{}
\providecommand*{\mciteBstWouldAddEndPuncttrue}
  {\def\EndOfBibitem{\unskip.}}
\providecommand*{\mciteBstWouldAddEndPunctfalse}
  {\let\EndOfBibitem\relax}
\providecommand*{\mciteSetBstMidEndSepPunct}[3]{}
\providecommand*{\mciteSetBstSublistLabelBeginEnd}[3]{}
\providecommand*{\EndOfBibitem}{}
\mciteSetBstSublistMode{f}
\mciteSetBstMaxWidthForm{subitem}
{(\emph{\alph{mcitesubitemcount}})}
\mciteSetBstSublistLabelBeginEnd{\mcitemaxwidthsubitemform\space}
{\relax}{\relax}

\bibitem[Berthier \emph{et~al.}(2005)Berthier, Biroli, Bouchaud, Cipelletti,
  {El Masri}, L'H\^{o}te, Ladieu, and Pierno]{Berthier2005a}
L.~Berthier, G.~Biroli, J.-P. Bouchaud, L.~Cipelletti, D.~{El Masri},
  D.~L'H\^{o}te, F.~Ladieu and M.~Pierno, \emph{Science (New York, N.Y.)},
  2005, \textbf{310}, 1797--800\relax
\mciteBstWouldAddEndPuncttrue
\mciteSetBstMidEndSepPunct{\mcitedefaultmidpunct}
{\mcitedefaultendpunct}{\mcitedefaultseppunct}\relax
\EndOfBibitem
\bibitem[Heussinger \emph{et~al.}(2010)Heussinger, Chaudhuri, and
  Barrat]{heussinger2010fluctuations}
C.~Heussinger, P.~Chaudhuri and J.-L. Barrat, \emph{Soft matter}, 2010,
  \textbf{6}, 3050--3058\relax
\mciteBstWouldAddEndPuncttrue
\mciteSetBstMidEndSepPunct{\mcitedefaultmidpunct}
{\mcitedefaultendpunct}{\mcitedefaultseppunct}\relax
\EndOfBibitem
\bibitem[Rodney \emph{et~al.}(2011)Rodney, Tanguy, and
  Vandembroucq]{Rodney2011Review}
D.~Rodney, A.~Tanguy and D.~Vandembroucq, \emph{Modelling and Simulation in
  Materials Science and Engineering}, 2011, \textbf{19}, 083001\relax
\mciteBstWouldAddEndPuncttrue
\mciteSetBstMidEndSepPunct{\mcitedefaultmidpunct}
{\mcitedefaultendpunct}{\mcitedefaultseppunct}\relax
\EndOfBibitem
\bibitem[Dahmen \emph{et~al.}(2011)Dahmen, Ben-Zion, and Uhl]{Dahmen2011}
K.~A. Dahmen, Y.~Ben-Zion and J.~T. Uhl, \emph{Nature Physics}, 2011,
  \textbf{7}, 554--557\relax
\mciteBstWouldAddEndPuncttrue
\mciteSetBstMidEndSepPunct{\mcitedefaultmidpunct}
{\mcitedefaultendpunct}{\mcitedefaultseppunct}\relax
\EndOfBibitem
\bibitem[Argon and Kuo(1979)]{Argon1979}
A.~Argon and H.~Kuo, \emph{Materials Science and Engineering}, 1979,
  \textbf{39}, 101--109\relax
\mciteBstWouldAddEndPuncttrue
\mciteSetBstMidEndSepPunct{\mcitedefaultmidpunct}
{\mcitedefaultendpunct}{\mcitedefaultseppunct}\relax
\EndOfBibitem
\bibitem[Lema\^{\i}tre and Caroli(2007)]{Lemaitre2007}
A.~Lema\^{\i}tre and C.~Caroli, \emph{Physical Review E}, 2007, \textbf{76},
  036104\relax
\mciteBstWouldAddEndPuncttrue
\mciteSetBstMidEndSepPunct{\mcitedefaultmidpunct}
{\mcitedefaultendpunct}{\mcitedefaultseppunct}\relax
\EndOfBibitem
\bibitem[Tsamados \emph{et~al.}(2008)Tsamados, Tanguy, L\'{e}onforte, and
  Barrat]{Tsamados2008}
M.~Tsamados, A.~Tanguy, F.~L\'{e}onforte and J.-L. Barrat, \emph{The European
  physical journal. E, Soft matter}, 2008, \textbf{26}, 283--93\relax
\mciteBstWouldAddEndPuncttrue
\mciteSetBstMidEndSepPunct{\mcitedefaultmidpunct}
{\mcitedefaultendpunct}{\mcitedefaultseppunct}\relax
\EndOfBibitem
\bibitem[Falk and Langer(1998)]{Falk1998}
M.~Falk and J.~Langer, \emph{Physical Review E}, 1998, \textbf{57}, 7192\relax
\mciteBstWouldAddEndPuncttrue
\mciteSetBstMidEndSepPunct{\mcitedefaultmidpunct}
{\mcitedefaultendpunct}{\mcitedefaultseppunct}\relax
\EndOfBibitem
\bibitem[Falk and Langer({2011})]{stz-review}
M.~L. Falk and J.~S. Langer, \emph{{ANNUAL REVIEW OF CONDENSED MATTER PHYSICS,
  VOL 2}}, {2011}, vol.~{2}, pp. {353--373}\relax
\mciteBstWouldAddEndPuncttrue
\mciteSetBstMidEndSepPunct{\mcitedefaultmidpunct}
{\mcitedefaultendpunct}{\mcitedefaultseppunct}\relax
\EndOfBibitem
\bibitem[Sollich \emph{et~al.}(1997)Sollich, Lequeux, H\'{e}braud, and
  Cates]{Sollich1997}
P.~Sollich, F.~Lequeux, P.~H\'{e}braud and M.~Cates, \emph{Physical Review
  Letters}, 1997, \textbf{78}, 2020--2023\relax
\mciteBstWouldAddEndPuncttrue
\mciteSetBstMidEndSepPunct{\mcitedefaultmidpunct}
{\mcitedefaultendpunct}{\mcitedefaultseppunct}\relax
\EndOfBibitem
\bibitem[Sollich(1998)]{Sollich1998}
P.~Sollich, \emph{Physical Review E}, 1998, \textbf{58}, 738\relax
\mciteBstWouldAddEndPuncttrue
\mciteSetBstMidEndSepPunct{\mcitedefaultmidpunct}
{\mcitedefaultendpunct}{\mcitedefaultseppunct}\relax
\EndOfBibitem
\bibitem[H\'{e}braud and Lequeux(1998)]{Hebraud1998}
P.~H\'{e}braud and F.~Lequeux, \emph{Physical Review Letters}, 1998,
  \textbf{81}, 2934--2937\relax
\mciteBstWouldAddEndPuncttrue
\mciteSetBstMidEndSepPunct{\mcitedefaultmidpunct}
{\mcitedefaultendpunct}{\mcitedefaultseppunct}\relax
\EndOfBibitem
\bibitem[Coussot \emph{et~al.}(2002)Coussot, Nguyen, Huynh, and
  Bonn]{Coussot2002}
P.~Coussot, Q.~Nguyen, H.~Huynh and D.~Bonn, \emph{Physical Review Letters},
  2002, \textbf{88}, 175501\relax
\mciteBstWouldAddEndPuncttrue
\mciteSetBstMidEndSepPunct{\mcitedefaultmidpunct}
{\mcitedefaultendpunct}{\mcitedefaultseppunct}\relax
\EndOfBibitem
\bibitem[Manning \emph{et~al.}(2007)Manning, Langer, and Carlson]{Manning2007}
M.~Manning, J.~Langer and J.~Carlson, \emph{Physical Review E}, 2007,
  \textbf{76}, 056106\relax
\mciteBstWouldAddEndPuncttrue
\mciteSetBstMidEndSepPunct{\mcitedefaultmidpunct}
{\mcitedefaultendpunct}{\mcitedefaultseppunct}\relax
\EndOfBibitem
\bibitem[Fielding \emph{et~al.}(2009)Fielding, Cates, and
  Sollich]{Fielding2009}
S.~M. Fielding, M.~E. Cates and P.~Sollich, \emph{Soft Matter}, 2009,
  \textbf{5}, 2378\relax
\mciteBstWouldAddEndPuncttrue
\mciteSetBstMidEndSepPunct{\mcitedefaultmidpunct}
{\mcitedefaultendpunct}{\mcitedefaultseppunct}\relax
\EndOfBibitem
\bibitem[Bocquet \emph{et~al.}(2009)Bocquet, Colin, and Ajdari]{Bocquet2009}
L.~Bocquet, A.~Colin and A.~Ajdari, \emph{Physical Review Letters}, 2009,
  \textbf{103}, 036001\relax
\mciteBstWouldAddEndPuncttrue
\mciteSetBstMidEndSepPunct{\mcitedefaultmidpunct}
{\mcitedefaultendpunct}{\mcitedefaultseppunct}\relax
\EndOfBibitem
\bibitem[Chen \emph{et~al.}(1991)Chen, Bak, and Obukhov]{Chen1991}
K.~Chen, P.~Bak and S.~Obukhov, \emph{Physical Review A}, 1991, \textbf{43},
  625--630\relax
\mciteBstWouldAddEndPuncttrue
\mciteSetBstMidEndSepPunct{\mcitedefaultmidpunct}
{\mcitedefaultendpunct}{\mcitedefaultseppunct}\relax
\EndOfBibitem
\bibitem[Bulatov and Argon(1994)]{Bulatov1994}
V.~V. Bulatov and A.~S. Argon, \emph{Modelling and Simulation in Materials
  Science and Engineering}, 1994, \textbf{2}, 167--184\relax
\mciteBstWouldAddEndPuncttrue
\mciteSetBstMidEndSepPunct{\mcitedefaultmidpunct}
{\mcitedefaultendpunct}{\mcitedefaultseppunct}\relax
\EndOfBibitem
\bibitem[Bulatov and Argon(1994)]{Bulatov1994a}
V.~V. Bulatov and A.~S. Argon, \emph{Modelling and Simulation in Materials
  Science and Engineering}, 1994, \textbf{2}, 185--202\relax
\mciteBstWouldAddEndPuncttrue
\mciteSetBstMidEndSepPunct{\mcitedefaultmidpunct}
{\mcitedefaultendpunct}{\mcitedefaultseppunct}\relax
\EndOfBibitem
\bibitem[Bulatov and Argon(1994)]{Bulatov1994b}
V.~V. Bulatov and A.~S. Argon, \emph{Modelling and Simulation in Materials
  Science and Engineering}, 1994, \textbf{2}, 203--222\relax
\mciteBstWouldAddEndPuncttrue
\mciteSetBstMidEndSepPunct{\mcitedefaultmidpunct}
{\mcitedefaultendpunct}{\mcitedefaultseppunct}\relax
\EndOfBibitem
\bibitem[Baret \emph{et~al.}(2002)Baret, Vandembroucq, and Roux]{Baret2002}
J.-C. Baret, D.~Vandembroucq and S.~Roux, \emph{Physical Review Letters}, 2002,
  \textbf{89}, 195506\relax
\mciteBstWouldAddEndPuncttrue
\mciteSetBstMidEndSepPunct{\mcitedefaultmidpunct}
{\mcitedefaultendpunct}{\mcitedefaultseppunct}\relax
\EndOfBibitem
\bibitem[Picard \emph{et~al.}(2005)Picard, Ajdari, Lequeux, and
  Bocquet]{Picard2005}
G.~Picard, A.~Ajdari, F.~Lequeux and L.~Bocquet, \emph{Physical Review E},
  2005, \textbf{71}, 010501\relax
\mciteBstWouldAddEndPuncttrue
\mciteSetBstMidEndSepPunct{\mcitedefaultmidpunct}
{\mcitedefaultendpunct}{\mcitedefaultseppunct}\relax
\EndOfBibitem
\bibitem[Homer and Schuh(2009)]{Homer2009}
E.~R. Homer and C.~A. Schuh, \emph{Acta Materialia}, 2009, \textbf{57},
  2823--2833\relax
\mciteBstWouldAddEndPuncttrue
\mciteSetBstMidEndSepPunct{\mcitedefaultmidpunct}
{\mcitedefaultendpunct}{\mcitedefaultseppunct}\relax
\EndOfBibitem
\bibitem[Martens \emph{et~al.}(2012)Martens, Bocquet, and Barrat]{Martens2012}
K.~Martens, L.~Bocquet and J.-L. Barrat, \emph{Soft Matter}, 2012,  4197\relax
\mciteBstWouldAddEndPuncttrue
\mciteSetBstMidEndSepPunct{\mcitedefaultmidpunct}
{\mcitedefaultendpunct}{\mcitedefaultseppunct}\relax
\EndOfBibitem
\bibitem[Martens \emph{et~al.}(2011)Martens, Bocquet, and Barrat]{Martens2011}
K.~Martens, L.~Bocquet and J.-L. Barrat, \emph{Physical Review Letters}, 2011,
  \textbf{106}, 156001\relax
\mciteBstWouldAddEndPuncttrue
\mciteSetBstMidEndSepPunct{\mcitedefaultmidpunct}
{\mcitedefaultendpunct}{\mcitedefaultseppunct}\relax
\EndOfBibitem
\bibitem[Nicolas and Barrat(2013)]{Nicolas2013}
A.~Nicolas and J.-L. Barrat, \emph{Physical Review Letters}, 2013,
  \textbf{110}, 138304\relax
\mciteBstWouldAddEndPuncttrue
\mciteSetBstMidEndSepPunct{\mcitedefaultmidpunct}
{\mcitedefaultendpunct}{\mcitedefaultseppunct}\relax
\EndOfBibitem
\bibitem[PRINCEN(1985)]{PRINCEN1985}
H.~PRINCEN, \emph{Journal of Colloid and Interface Science}, 1985,
  \textbf{105}, 150--171\relax
\mciteBstWouldAddEndPuncttrue
\mciteSetBstMidEndSepPunct{\mcitedefaultmidpunct}
{\mcitedefaultendpunct}{\mcitedefaultseppunct}\relax
\EndOfBibitem
\bibitem[Maloney and Lema\^{\i}tre(2006)]{Maloney2006}
C.~Maloney and A.~Lema\^{\i}tre, \emph{Physical Review E}, 2006, \textbf{74},
  016118\relax
\mciteBstWouldAddEndPuncttrue
\mciteSetBstMidEndSepPunct{\mcitedefaultmidpunct}
{\mcitedefaultendpunct}{\mcitedefaultseppunct}\relax
\EndOfBibitem
\bibitem[Amon \emph{et~al.}(2012)Amon, Nguyen, Bruand, Crassous, and
  Cl\'{e}ment]{Amon2012}
A.~Amon, V.~Nguyen, A.~Bruand, J.~Crassous and E.~Cl\'{e}ment, \emph{Physical
  Review Letters}, 2012, \textbf{108}, 135502\relax
\mciteBstWouldAddEndPuncttrue
\mciteSetBstMidEndSepPunct{\mcitedefaultmidpunct}
{\mcitedefaultendpunct}{\mcitedefaultseppunct}\relax
\EndOfBibitem
\bibitem[{Le Merrer} \emph{et~al.}(2012){Le Merrer}, Cohen-Addad, and
  H\"{o}hler]{LeMerrer2012}
M.~{Le Merrer}, S.~Cohen-Addad and R.~H\"{o}hler, \emph{Physical Review
  Letters}, 2012, \textbf{108}, 188301\relax
\mciteBstWouldAddEndPuncttrue
\mciteSetBstMidEndSepPunct{\mcitedefaultmidpunct}
{\mcitedefaultendpunct}{\mcitedefaultseppunct}\relax
\EndOfBibitem
\bibitem[Bouchaud and Pitard({2001})]{Bouchaud2001}
J.~Bouchaud and E.~Pitard, \emph{{The European Physical Journal E}}, {2001},
  \textbf{{6}}, {231--236}\relax
\mciteBstWouldAddEndPuncttrue
\mciteSetBstMidEndSepPunct{\mcitedefaultmidpunct}
{\mcitedefaultendpunct}{\mcitedefaultseppunct}\relax
\EndOfBibitem
\bibitem[Cloitre \emph{et~al.}(2003)Cloitre, Borrega, Monti, and
  Leibler]{Cloitre2003}
M.~Cloitre, R.~Borrega, F.~Monti and L.~Leibler, \emph{Physical Review
  Letters}, 2003, \textbf{90}, 068303\relax
\mciteBstWouldAddEndPuncttrue
\mciteSetBstMidEndSepPunct{\mcitedefaultmidpunct}
{\mcitedefaultendpunct}{\mcitedefaultseppunct}\relax
\EndOfBibitem
\bibitem[Picard \emph{et~al.}(2004)Picard, Ajdari, Lequeux, and
  Bocquet]{Picard2004}
G.~Picard, A.~Ajdari, F.~Lequeux and L.~Bocquet, \emph{The European physical
  journal. E, Soft matter}, 2004, \textbf{15}, 371--81\relax
\mciteBstWouldAddEndPuncttrue
\mciteSetBstMidEndSepPunct{\mcitedefaultmidpunct}
{\mcitedefaultendpunct}{\mcitedefaultseppunct}\relax
\EndOfBibitem
\bibitem[Leonforte \emph{et~al.}(2005)Leonforte, Boissi\`{e}re, Tanguy,
  Wittmer, and Barrat]{Leonforte2005}
F.~Leonforte, R.~Boissi\`{e}re, A.~Tanguy, J.~Wittmer and J.-L. Barrat,
  \emph{Physical Review B}, 2005, \textbf{72}, 224206\relax
\mciteBstWouldAddEndPuncttrue
\mciteSetBstMidEndSepPunct{\mcitedefaultmidpunct}
{\mcitedefaultendpunct}{\mcitedefaultseppunct}\relax
\EndOfBibitem
\bibitem[Schall \emph{et~al.}(2007)Schall, Weitz, and Spaepen]{Schall2007}
P.~Schall, D.~A. Weitz and F.~Spaepen, \emph{Science (New York, N.Y.)}, 2007,
  \textbf{318}, 1895--9\relax
\mciteBstWouldAddEndPuncttrue
\mciteSetBstMidEndSepPunct{\mcitedefaultmidpunct}
{\mcitedefaultendpunct}{\mcitedefaultseppunct}\relax
\EndOfBibitem
\bibitem[Bokeloh \emph{et~al.}(2011)Bokeloh, Divinski, Reglitz, and
  Wilde]{Bokeloh2011}
J.~Bokeloh, S.~V. Divinski, G.~Reglitz and G.~Wilde, \emph{Physical Review
  Letters}, 2011, \textbf{107}, 235503\relax
\mciteBstWouldAddEndPuncttrue
\mciteSetBstMidEndSepPunct{\mcitedefaultmidpunct}
{\mcitedefaultendpunct}{\mcitedefaultseppunct}\relax
\EndOfBibitem
\bibitem[Besseling \emph{et~al.}(2010)Besseling, Isa, Ballesta, Petekidis,
  Cates, and Poon]{Besseling2010}
R.~Besseling, L.~Isa, P.~Ballesta, G.~Petekidis, M.~Cates and W.~Poon,
  \emph{Physical Review Letters}, 2010, \textbf{105}, 268301\relax
\mciteBstWouldAddEndPuncttrue
\mciteSetBstMidEndSepPunct{\mcitedefaultmidpunct}
{\mcitedefaultendpunct}{\mcitedefaultseppunct}\relax
\EndOfBibitem
\bibitem[Talamali \emph{et~al.}(2012)Talamali, Pet{\"a}j{\"a}, Vandembroucq,
  and Roux]{Talamali2012}
M.~Talamali, V.~Pet{\"a}j{\"a}, D.~Vandembroucq and S.~Roux, \emph{Comptes
  Rendus de M{\'e}canique}, 2012, \textbf{340}, 275\relax
\mciteBstWouldAddEndPuncttrue
\mciteSetBstMidEndSepPunct{\mcitedefaultmidpunct}
{\mcitedefaultendpunct}{\mcitedefaultseppunct}\relax
\EndOfBibitem
\bibitem[Goyon \emph{et~al.}(2008)Goyon, Colin, Ovarlez, Ajdari, and
  Bocquet]{Goyon2008a}
J.~Goyon, A.~Colin, G.~Ovarlez, A.~Ajdari and L.~Bocquet, \emph{Nature}, 2008,
  \textbf{454}, 84--7\relax
\mciteBstWouldAddEndPuncttrue
\mciteSetBstMidEndSepPunct{\mcitedefaultmidpunct}
{\mcitedefaultendpunct}{\mcitedefaultseppunct}\relax
\EndOfBibitem
\bibitem[Chaudhuri \emph{et~al.}(2012)Chaudhuri, Mansard, Colin, and
  Bocquet]{Chaudhuri2012}
P.~Chaudhuri, V.~Mansard, A.~Colin and L.~Bocquet, \emph{Physical Review
  Letters}, 2012, \textbf{109}, 036001\relax
\mciteBstWouldAddEndPuncttrue
\mciteSetBstMidEndSepPunct{\mcitedefaultmidpunct}
{\mcitedefaultendpunct}{\mcitedefaultseppunct}\relax
\EndOfBibitem
\bibitem[Pouliquen \emph{et~al.}(2001)Pouliquen, Forterre, and {Le
  Dizes}]{Pouliquen2001}
O.~Pouliquen, Y.~Forterre and S.~{Le Dizes}, \emph{Advances in Complex
  Systems}, 2001, \textbf{04}, 441--450\relax
\mciteBstWouldAddEndPuncttrue
\mciteSetBstMidEndSepPunct{\mcitedefaultmidpunct}
{\mcitedefaultendpunct}{\mcitedefaultseppunct}\relax
\EndOfBibitem
\bibitem[Hartley and Behringer(2003)]{Hartley2003}
R.~R. Hartley and R.~P. Behringer, \emph{Nature}, 2003, \textbf{421},
  928--31\relax
\mciteBstWouldAddEndPuncttrue
\mciteSetBstMidEndSepPunct{\mcitedefaultmidpunct}
{\mcitedefaultendpunct}{\mcitedefaultseppunct}\relax
\EndOfBibitem
\bibitem[Amon \emph{et~al.}(2012)Amon, Bertoni, and Crassous]{Amon2012a}
A.~Amon, R.~Bertoni and J.~Crassous, 2012\relax
\mciteBstWouldAddEndPuncttrue
\mciteSetBstMidEndSepPunct{\mcitedefaultmidpunct}
{\mcitedefaultendpunct}{\mcitedefaultseppunct}\relax
\EndOfBibitem
\bibitem[Goyon \emph{et~al.}(2010)Goyon, Colin, and Bocquet]{Goyon2010}
J.~Goyon, A.~Colin and L.~Bocquet, \emph{Soft Matter}, 2010, \textbf{6},
  2668\relax
\mciteBstWouldAddEndPuncttrue
\mciteSetBstMidEndSepPunct{\mcitedefaultmidpunct}
{\mcitedefaultendpunct}{\mcitedefaultseppunct}\relax
\EndOfBibitem
\bibitem[Jop \emph{et~al.}(2012)Jop, Mansard, Chaudhuri, Bocquet, and
  Colin]{Jop2012}
P.~Jop, V.~Mansard, P.~Chaudhuri, L.~Bocquet and A.~Colin, \emph{Physical
  Review Letters}, 2012, \textbf{108}, 148301\relax
\mciteBstWouldAddEndPuncttrue
\mciteSetBstMidEndSepPunct{\mcitedefaultmidpunct}
{\mcitedefaultendpunct}{\mcitedefaultseppunct}\relax
\EndOfBibitem
\bibitem[Geraud \emph{et~al.}(2013)Geraud, Bocquet, and Barentin]{Geraud2013}
B.~Geraud, L.~Bocquet and C.~Barentin, \emph{The European physical journal. E,
  Soft matter}, 2013, \textbf{36}, 9845\relax
\mciteBstWouldAddEndPuncttrue
\mciteSetBstMidEndSepPunct{\mcitedefaultmidpunct}
{\mcitedefaultendpunct}{\mcitedefaultseppunct}\relax
\EndOfBibitem
\bibitem[Isa \emph{et~al.}(2007)Isa, Besseling, and Poon]{Isa2007}
L.~Isa, R.~Besseling and W.~C.~K. Poon, \emph{Physical Review Letters}, 2007,
  \textbf{98}, 198305\relax
\mciteBstWouldAddEndPuncttrue
\mciteSetBstMidEndSepPunct{\mcitedefaultmidpunct}
{\mcitedefaultendpunct}{\mcitedefaultseppunct}\relax
\EndOfBibitem
\bibitem[Isa \emph{et~al.}(2009)Isa, Besseling, Morozov, and Poon]{Isa2009}
L.~Isa, R.~Besseling, A.~Morozov and W.~Poon, \emph{Physical Review Letters},
  2009, \textbf{102}, 058302\relax
\mciteBstWouldAddEndPuncttrue
\mciteSetBstMidEndSepPunct{\mcitedefaultmidpunct}
{\mcitedefaultendpunct}{\mcitedefaultseppunct}\relax
\EndOfBibitem
\bibitem[Pouliquen and Gutfraind(1996)]{Pouliquen1996}
O.~Pouliquen and R.~Gutfraind, \emph{Physical Review E}, 1996, \textbf{53},
  552\relax
\mciteBstWouldAddEndPuncttrue
\mciteSetBstMidEndSepPunct{\mcitedefaultmidpunct}
{\mcitedefaultendpunct}{\mcitedefaultseppunct}\relax
\EndOfBibitem
\bibitem[Gutfraind and Pouliquen(1996)]{Gutfraind1996}
R.~Gutfraind and O.~Pouliquen, \emph{Mechanics of materials}, 1996,
  \textbf{24}, 273--285\relax
\mciteBstWouldAddEndPuncttrue
\mciteSetBstMidEndSepPunct{\mcitedefaultmidpunct}
{\mcitedefaultendpunct}{\mcitedefaultseppunct}\relax
\EndOfBibitem
\bibitem[Salmon \emph{et~al.}(2003)Salmon, B{\'e}cu, Manneville, and
  Colin]{Salmon2003Emulsions}
J.-B. Salmon, L.~B{\'e}cu, S.~Manneville and A.~Colin, \emph{The European
  Physical Journal E}, 2003, \textbf{10}, 209--221\relax
\mciteBstWouldAddEndPuncttrue
\mciteSetBstMidEndSepPunct{\mcitedefaultmidpunct}
{\mcitedefaultendpunct}{\mcitedefaultseppunct}\relax
\EndOfBibitem
\bibitem[Kamrin and Koval(2012)]{Kamrin2012}
K.~Kamrin and G.~Koval, \emph{Physical Review Letters}, 2012, \textbf{108},
  178301\relax
\mciteBstWouldAddEndPuncttrue
\mciteSetBstMidEndSepPunct{\mcitedefaultmidpunct}
{\mcitedefaultendpunct}{\mcitedefaultseppunct}\relax
\EndOfBibitem
\bibitem[Ovarlez \emph{et~al.}(2008)Ovarlez, Rodts, Ragouilliaux, Coussot,
  Goyon, and Colin]{Ovarlez2008}
G.~Ovarlez, S.~Rodts, A.~Ragouilliaux, P.~Coussot, J.~Goyon and A.~Colin,
  \emph{Physical Review E}, 2008, \textbf{78}, 036307\relax
\mciteBstWouldAddEndPuncttrue
\mciteSetBstMidEndSepPunct{\mcitedefaultmidpunct}
{\mcitedefaultendpunct}{\mcitedefaultseppunct}\relax
\EndOfBibitem
\bibitem[Masselon and Colin(2010)]{Masselon2010}
C.~Masselon and A.~Colin, \emph{Physical Review E}, 2010, \textbf{81},
  021502\relax
\mciteBstWouldAddEndPuncttrue
\mciteSetBstMidEndSepPunct{\mcitedefaultmidpunct}
{\mcitedefaultendpunct}{\mcitedefaultseppunct}\relax
\EndOfBibitem
\bibitem[B\'{e}cu \emph{et~al.}(2004)B\'{e}cu, Manneville, and Colin]{Becu2004}
L.~B\'{e}cu, S.~Manneville and A.~Colin, \emph{Physical Review Letters}, 2004,
  \textbf{93}, 018301\relax
\mciteBstWouldAddEndPuncttrue
\mciteSetBstMidEndSepPunct{\mcitedefaultmidpunct}
{\mcitedefaultendpunct}{\mcitedefaultseppunct}\relax
\EndOfBibitem
\bibitem[Gibaud \emph{et~al.}(2008)Gibaud, Barentin, and
  Manneville]{Gibaud2008}
T.~Gibaud, C.~Barentin and S.~Manneville, \emph{Physical Review Letters}, 2008,
  \textbf{101}, 258302\relax
\mciteBstWouldAddEndPuncttrue
\mciteSetBstMidEndSepPunct{\mcitedefaultmidpunct}
{\mcitedefaultendpunct}{\mcitedefaultseppunct}\relax
\EndOfBibitem
\bibitem[Ballesta \emph{et~al.}(2008)Ballesta, Besseling, Isa, Petekidis, and
  Poon]{Ballesta2008}
P.~Ballesta, R.~Besseling, L.~Isa, G.~Petekidis and W.~Poon, \emph{Physical
  Review Letters}, 2008, \textbf{101}, 258301\relax
\mciteBstWouldAddEndPuncttrue
\mciteSetBstMidEndSepPunct{\mcitedefaultmidpunct}
{\mcitedefaultendpunct}{\mcitedefaultseppunct}\relax
\EndOfBibitem
\bibitem[Mansard(2012)]{Mansard2012PhD}
V.~Mansard, \emph{PhD thesis}, Universit\'e de Bordeaux I, 2012\relax
\mciteBstWouldAddEndPuncttrue
\mciteSetBstMidEndSepPunct{\mcitedefaultmidpunct}
{\mcitedefaultendpunct}{\mcitedefaultseppunct}\relax
\EndOfBibitem
\bibitem[Pagac \emph{et~al.}(1996)Pagac, Tilton, and Prieve]{Pagac1996}
E.~Pagac, R.~Tilton and D.~Prieve, \emph{Chemical engineering communications},
  1996, \textbf{148}, 105--122\relax
\mciteBstWouldAddEndPuncttrue
\mciteSetBstMidEndSepPunct{\mcitedefaultmidpunct}
{\mcitedefaultendpunct}{\mcitedefaultseppunct}\relax
\EndOfBibitem
\bibitem[Yoshimura and Prud'homme(1988)]{yoshimura1988wall}
A.~Yoshimura and R.~K. Prud'homme, \emph{Journal of Rheology}, 1988,
  \textbf{32}, 53\relax
\mciteBstWouldAddEndPuncttrue
\mciteSetBstMidEndSepPunct{\mcitedefaultmidpunct}
{\mcitedefaultendpunct}{\mcitedefaultseppunct}\relax
\EndOfBibitem
\bibitem[Barnes(1995)]{Barnes1995}
H.~A. Barnes, \emph{Journal of Non-Newtonian Fluid Mechanics}, 1995,
  \textbf{56}, 221--251\relax
\mciteBstWouldAddEndPuncttrue
\mciteSetBstMidEndSepPunct{\mcitedefaultmidpunct}
{\mcitedefaultendpunct}{\mcitedefaultseppunct}\relax
\EndOfBibitem
\bibitem[Franco \emph{et~al.}(1998)Franco, Gallegos, and Barnes]{Franco1998}
J.~Franco, C.~Gallegos and H.~Barnes, \emph{Journal of Food Engineering}, 1998,
  \textbf{36}, 89--102\relax
\mciteBstWouldAddEndPuncttrue
\mciteSetBstMidEndSepPunct{\mcitedefaultmidpunct}
{\mcitedefaultendpunct}{\mcitedefaultseppunct}\relax
\EndOfBibitem
\bibitem[Meeker \emph{et~al.}(2004)Meeker, Bonnecaze, and Cloitre]{Meeker2004}
S.~Meeker, R.~Bonnecaze and M.~Cloitre, \emph{Physical Review Letters}, 2004,
  \textbf{92}, 198302\relax
\mciteBstWouldAddEndPuncttrue
\mciteSetBstMidEndSepPunct{\mcitedefaultmidpunct}
{\mcitedefaultendpunct}{\mcitedefaultseppunct}\relax
\EndOfBibitem
\bibitem[Meeker \emph{et~al.}(2004)Meeker, Bonnecaze, and
  Cloitre]{Meeker2004long}
S.~P. Meeker, R.~T. Bonnecaze and M.~Cloitre, \emph{Journal of Rheology}, 2004,
  \textbf{48}, 1295\relax
\mciteBstWouldAddEndPuncttrue
\mciteSetBstMidEndSepPunct{\mcitedefaultmidpunct}
{\mcitedefaultendpunct}{\mcitedefaultseppunct}\relax
\EndOfBibitem
\bibitem[Salmon \emph{et~al.}(2003)Salmon, Colin, and Manneville]{Salmon2003}
J.-B. Salmon, A.~Colin and S.~Manneville, \emph{Physical Review Letters}, 2003,
  \textbf{90}, 228303\relax
\mciteBstWouldAddEndPuncttrue
\mciteSetBstMidEndSepPunct{\mcitedefaultmidpunct}
{\mcitedefaultendpunct}{\mcitedefaultseppunct}\relax
\EndOfBibitem
\bibitem[B{\'e}cu \emph{et~al.}(2005)B{\'e}cu, Grondin, Colin, and
  Manneville]{Becu2005}
L.~B{\'e}cu, P.~Grondin, A.~Colin and S.~Manneville, \emph{Colloids and
  Surfaces A: Physicochemical and Engineering Aspects}, 2005, \textbf{263},
  146--152\relax
\mciteBstWouldAddEndPuncttrue
\mciteSetBstMidEndSepPunct{\mcitedefaultmidpunct}
{\mcitedefaultendpunct}{\mcitedefaultseppunct}\relax
\EndOfBibitem
\bibitem[Mason \emph{et~al.}(1996)Mason, Bibette, and Weitz]{Mason1996JColl}
T.~Mason, J.~Bibette and D.~Weitz, \emph{Journal of Colloid and Interface
  Science}, 1996, \textbf{179}, 439--448\relax
\mciteBstWouldAddEndPuncttrue
\mciteSetBstMidEndSepPunct{\mcitedefaultmidpunct}
{\mcitedefaultendpunct}{\mcitedefaultseppunct}\relax
\EndOfBibitem
\bibitem[Sanchez \emph{et~al.}(2001)Sanchez, Valencia, Franco, and
  Gallegos]{Sanchez2001}
M.~Sanchez, C.~Valencia, J.~Franco and C.~Gallegos, \emph{Journal of colloid
  and interface science}, 2001, \textbf{241}, 226--232\relax
\mciteBstWouldAddEndPuncttrue
\mciteSetBstMidEndSepPunct{\mcitedefaultmidpunct}
{\mcitedefaultendpunct}{\mcitedefaultseppunct}\relax
\EndOfBibitem
\bibitem[Meeten(2004)]{Meeten2004}
G.~H. Meeten, \emph{Journal of non-newtonian fluid mechanics}, 2004,
  \textbf{124}, 51--60\relax
\mciteBstWouldAddEndPuncttrue
\mciteSetBstMidEndSepPunct{\mcitedefaultmidpunct}
{\mcitedefaultendpunct}{\mcitedefaultseppunct}\relax
\EndOfBibitem
\bibitem[Goyon(2008)]{Goyon2008PhD}
J.~Goyon, \emph{PhD thesis}, Universit\'e de Bordeaux I, 2008\relax
\mciteBstWouldAddEndPuncttrue
\mciteSetBstMidEndSepPunct{\mcitedefaultmidpunct}
{\mcitedefaultendpunct}{\mcitedefaultseppunct}\relax
\EndOfBibitem
\bibitem[Divoux \emph{et~al.}(2011)Divoux, Barentin, and
  Manneville]{Divoux2011Overshoot}
T.~Divoux, C.~Barentin and S.~Manneville, \emph{Soft Matter}, 2011, \textbf{7},
  9335--9349\relax
\mciteBstWouldAddEndPuncttrue
\mciteSetBstMidEndSepPunct{\mcitedefaultmidpunct}
{\mcitedefaultendpunct}{\mcitedefaultseppunct}\relax
\EndOfBibitem
\bibitem[Divoux \emph{et~al.}(2011)Divoux, Barentin, and
  Manneville]{Divoux2011Herschel}
T.~Divoux, C.~Barentin and S.~Manneville, \emph{Soft Matter}, 2011, \textbf{7},
  8409--8418\relax
\mciteBstWouldAddEndPuncttrue
\mciteSetBstMidEndSepPunct{\mcitedefaultmidpunct}
{\mcitedefaultendpunct}{\mcitedefaultseppunct}\relax
\EndOfBibitem
\bibitem[Seth \emph{et~al.}(2012)Seth, Locatelli-Champagne, Monti, Bonnecaze,
  and Cloitre]{Seth2012}
J.~R. Seth, C.~Locatelli-Champagne, F.~Monti, R.~T. Bonnecaze and M.~Cloitre,
  \emph{Soft Matter}, 2012, \textbf{8}, 140\relax
\mciteBstWouldAddEndPuncttrue
\mciteSetBstMidEndSepPunct{\mcitedefaultmidpunct}
{\mcitedefaultendpunct}{\mcitedefaultseppunct}\relax
\EndOfBibitem
\bibitem[Seth \emph{et~al.}(2008)Seth, Cloitre, and Bonnecaze]{Seth2008}
J.~R. Seth, M.~Cloitre and R.~T. Bonnecaze, \emph{Journal of Rheology}, 2008,
  \textbf{52}, 1241\relax
\mciteBstWouldAddEndPuncttrue
\mciteSetBstMidEndSepPunct{\mcitedefaultmidpunct}
{\mcitedefaultendpunct}{\mcitedefaultseppunct}\relax
\EndOfBibitem
\bibitem[Besseling \emph{et~al.}(2009)Besseling, Isa, Weeks, and
  Poon]{Besseling2009}
R.~Besseling, L.~Isa, E.~R. Weeks and W.~C. Poon, \emph{Advances in colloid and
  interface science}, 2009, \textbf{146}, 1--17\relax
\mciteBstWouldAddEndPuncttrue
\mciteSetBstMidEndSepPunct{\mcitedefaultmidpunct}
{\mcitedefaultendpunct}{\mcitedefaultseppunct}\relax
\EndOfBibitem
\bibitem[Ballesta \emph{et~al.}(2012)Ballesta, Petekidis, Isa, Poon, and
  Besseling]{Ballesta2012}
P.~Ballesta, G.~Petekidis, L.~Isa, W.~C.~K. Poon and R.~Besseling,
  \emph{Journal of Rheology}, 2012, \textbf{56}, 1005\relax
\mciteBstWouldAddEndPuncttrue
\mciteSetBstMidEndSepPunct{\mcitedefaultmidpunct}
{\mcitedefaultendpunct}{\mcitedefaultseppunct}\relax
\EndOfBibitem
\bibitem[Gradshteyn and Ryzhik(1994)]{Gradshteyn1994}
I.~Gradshteyn and I.~Ryzhik, \emph{Tables of Integrals, Series and Products,
  (edited by A. Jeffrey)}, 1994\relax
\mciteBstWouldAddEndPuncttrue
\mciteSetBstMidEndSepPunct{\mcitedefaultmidpunct}
{\mcitedefaultendpunct}{\mcitedefaultseppunct}\relax
\EndOfBibitem
\end{mcitethebibliography}

\appendix

\appendixpage

\section{Derivation of the correction terms to the propagator
for a system bounded by walls\label{sec:Appendix_Derivation_Correction_terms}}

The system covers the domain $\left(x,y\right)\in\left[0,L_{x}\right]\times\left[-L_{y},L_{y}\right]$
and is periodically replicated throughout space. The region $y\in\left[0,L_{y}\right]$,
bounded by walls at $y=0$ and $y=L_{y}$ represents the \emph{real} system,
whereas the other half is a fictitious region introduced for the calculations.

For any plastic event $\boldsymbol{\epsilon}^{pl}=\left(\epsilon_{xx}^{pl},\epsilon_{xy}^{pl}\right)^{T}$
occurring at position $\left(x,y\right)$ in the real half, a 'symmetric'
plastic event $\boldsymbol{\epsilon}^{pl\,\prime}=\left(\epsilon_{xx}^{pl},-\epsilon_{xy}^{pl}\right)^{T}$
is created at location $\left(x,-y\right)$ in the fictitious region.
For symmetry reasons, the\emph{ y}-component of the velocity field
is thereby cancelled on lines $y=0$ and $y=L_{y}$ (bear in mind that
the $2L_{y}$-wide system is periodically replicated).

Let us now introduce forces $f_{x}^{\left(y=0\right)}$ and $f_{x}^{\left(y=L_{y}\right)}$
along the \emph{x}-direction at the bottom $\left(y=0\right)$ and
top $\left(y=L_{y}\right)$ walls, respectively, to cancel the \emph{x}-components.
The Fourier transform of the force field reads: 
\[
f_{x}(m,n)=f_{x}^{\left(y=0\right)}(m)+\left(-1\right)^{n}\, f_{x}^{\left(y=L_{y}\right)}(m)
\]
Note that we have simplified
notations by using the shorthand $g\left(m,n\right)$ for $\hat{g}\left(p_{m},q_{n}\right)$,
for any function $g$, where $p_{m}\equiv\frac{2\pi}{L_{x}}$ and $q_{n}\equiv\frac{2\pi}{2L_{y}}$
are the Fourier wavenumbers.

With these forces, the Fourier-transformed displacement field turns into:
\begin{eqnarray}
u^{(1)}\left(m,n\right) & = & \boldsymbol{\mathcal{G}^{\infty}}\left(m,n\right)\cdot\left(\hat{\boldsymbol{\epsilon}}^{pl}\left(m,n\right)+\hat{\boldsymbol{\epsilon}}^{pl\,\prime}\left(m,n\right)\right)+\boldsymbol{\mathcal{O}}\left(m,n\right)\cdot f_{x}\left(m,n\right)\label{eq:corrected_induced_velocity}\\
 & \equiv & u^{\star\infty}\left(m,n\right)+u^{corr}\left(m,n\right),
\end{eqnarray}
where $\hat{u}^{corr}$ is the contribution from the wall forces and
$\hat{\mathcal{O}}$ is the Oseen-Burgers tensor introduced in Eq.
\ref{eq:Oseen_Burgers_tensor}. The star in {$\hat{u}^{\star\infty}$
only indicates that this symbol represents the velocity field induced
by both the real plastic event and its 'symmetric' counterpart.}

Remarking that the condition of zero velocity at the bottom and top
walls reads, in terms of Fourier components, 
\[
\forall m,\ \sum_{n}u^{(1)}(m,n)=0
\] 
\[
\text{and }\forall m,\ \sum_{n}(-1)^{n}u^{(1)}(m,n)=0,
\] respectively, we
obtain two equations on the $f_{x}$ after insertion from Eq.\ref{eq:corrected_induced_velocity}.
 Adding and subtracting  these
equations yields, for any \emph{m}:

\[
\sum_{n\in O}u_{x}^{\star\infty}(m,n)+\mathcal{O}(m,n)\cdot\left(\left(\hat{f}_{x}^{\left(y=0\right)}-\hat{f}_{x}^{\left(y=L_{y}\right)}\right)(m)\right)=0
\]

\[
\sum_{n\in E}u_{x}^{\star\infty}(m,n)+\mathcal{O}(m,n)\cdot\left(\left(\hat{f}_{x}^{\left(y=0\right)}+\hat{f}_{x}^{\left(y=L_{y}\right)}\right)(m)\right)=0
\]
where $O\equiv2\mathbb{Z}+1$ is the set of odd integers, and $E\equiv2\mathbb{Z}$
is the set of even integers.

The solution of this linear system of equations is: 
\begin{equation}
f(m\neq0,n\in\delta)=\frac{-\mu}{e_{\delta}(m)}\sum_{n^{\prime}\in\delta}u_{x}^{\star\infty}(m,n^{\prime}),\label{eq:f_corr}
\end{equation}
where the symbol $\delta$ stands for either \emph{E }(even \emph{n}'s)
or \emph{O }(odd \emph{n}'s). The expressions for $m=0$ are written
separately: 
\[
f(0,n\in2\mathbb{Z})=0
\]

\[
f(0,n\in O)=\frac{-4\mu}{L_{y}^{2}}\sum_{n^{\prime}\in O}u_{x}^{\star\infty}(m,n^{\prime}).
\]

In Eq.\ref{eq:f_corr}, we have introduced auxiliary functions $e_{E}\left(m\right)$
and $e_{O}\left(m\right)$, which satisfy%
\footnote{The analytical calculations leading to the second part of the equality
involve the decomposition into simple elements and the use of well
established summation results\cite{Gradshteyn1994}.%
}: 
\[
e(m)\equiv\sum_{n\in\mathbb{Z}}\frac{q_{n}^{2}}{\left(p_{m}^{2}+q_{n}^{2}\right)^{2}}=\frac{L_{y}^{2}}{2\pi}\left[\frac{-\pi}{\sinh^{2}\left(\nicefrac{2\pi L_{y}m}{L_{x}}\right)}+\frac{L_{x}}{2mL_{y}}\frac{1}{\tanh\left(\nicefrac{2\pi L_{y}m}{L_{x}}\right)}\right]
\]
 
\[
e_{E}(m)\equiv\sum_{n\in E}\frac{q_{n}^{2}}{\left(p_{m}^{2}+q_{n}^{2}\right)^{2}}=\nicefrac{1}{4}\, e(\nicefrac{m}{2})
\]

\[
e_{O}(m)\equiv\sum_{n\in O}\frac{q_{n}^{2}}{\left(p_{m}^{2}+q_{n}^{2}\right)^{2}}=e(m)-\nicefrac{1}{4}\, e(\nicefrac{m}{2})
\]
Now, the infinite summation in Eq. \ref{eq:f_corr} needs to be calculated.
For a single plastic event located at $\left(x_{ev},y_{ev}\right)$,
that is, $\boldsymbol{\hat{\epsilon}^{pl}}\left(m,n\right)=e^{-ip_{m}x_{ev}}e^{-ip_{m}x_{ev}}\left(\epsilon_{xx}^{pl},\epsilon_{xy}^{pl}\right)^{T}$,
the use of the expression for $\hat{u_{x}}^{\star\infty}$ leads to:
\begin{equation}
\sum_{n^{\prime}\in \delta}\hat{u_{x}}^{\star\infty}(m,n^{\prime})=4e^{-ip_{m}x_{ev}}\left[\epsilon_{xy}^{pl}\left(p_{m}^{2}\frac{L_{y}^{3}}{\pi^{3}}j_{\delta}(X)-\frac{L_{y}}{\pi}k_{\delta}(X)\right)-2i\epsilon_{xx}^{pl}p_{m}\frac{L_{y}^{2}}{\pi^{2}}s_{\delta}(X)\right],\label{eq:somme_u_inf}
\end{equation}
where the $\delta$-subscript stands for either \emph{E }or \emph{O,
}and $X\equiv\left(x,\alpha\right)\equiv\left(\frac{\pi y_{ev}}{L_{y}},\frac{p_{m}L_{y}}{\pi}\right)$.

Inserting Eq.\ref{eq:somme_u_inf} into Eq.\ref{eq:f_corr}, summing
the plastic activity of all lines \emph{y}, i.e.%
\footnote{The $+0.5$ term comes from the fact that the \emph{y}-coordinate
of a block (of unit size) is evaluated at its centre.%
}, $y=0.5,\ldots,L_{y}-0.5\ \left(L_{y}\in\mathbb{N}^{\star}\right)$ in
the discretised version, and Fourier transforming the results along
direction \emph{x }via the operator $\mathcal{F}_{x}$, defined by
$\mathcal{F}_{x}\sigma=\nicefrac{1}{L_{x}}\int\sigma(x)e^{-ip_{m}x}dx$,
one finally arrives at~: {\small\[
\underline{\hat{u}}^{corr}(m,n\in\delta)=\left(\begin{array}{c}
\begin{array}{cc}
\equiv\zeta_{\delta}(X) & \equiv\xi_{\delta}(X)\\
\frac{-4q_{n}^{2}}{4\mu q^{4}}\cdot\Big[{\displaystyle \sum_{y}}\overbrace{\left(\frac{p_{m}^{2}L_{y}^{2}}{e_{\delta}(m)\pi^{3}}j_{\delta}(X)-\frac{1}{\pi}k_{\delta}(X)\right)}\mathcal{F}_{x}\sigma_{xy}^{pl}(m,y) & -2i{\displaystyle \sum_{y}}\overbrace{\left(\frac{p_{m}L_{y}}{e_{\delta}(m)\pi^{2}}s_{\delta}(X)\right)}\mathcal{F}_{x}\sigma_{xx}^{pl}(m,y)\Big]\\
\frac{4p_{m}q_{n}}{4\mu q^{4}}\Big[{\displaystyle \sum_{y}{\textstyle \left(\frac{p_{m}^{2}L_{y}^{2}}{e_{\delta}(m)\pi^{3}}j_{\delta}(X)-\frac{1}{\pi}k_{\delta}(X)\right)}}\mathcal{F}_{x}\sigma_{xy}^{pl}(m,y) & -2i{\displaystyle \sum_{y}}\frac{p_{m}L_{y}}{e_{\delta}(m)\pi^{2}}s_{\delta}(X)\mathcal{F}_{x}\sigma_{xx}^{pl}(m,y)\Big]
\end{array}\end{array}\right),
\]
}
where new summations appear and can be expressed analytically via
a decomposition into simple elements and the use of known summation
formulae\cite{Gradshteyn1994}:

\begin{alignat*}{1}
j(x,\alpha)\equiv & \sum_{k=-\infty}^{+\infty}\frac{k\sin\left(kx\right)}{\left(k^{2}+\alpha^{2}\right)^{2}}=\frac{\pi}{2\alpha^{2}}\frac{\sinh\left(\alpha\left(\pi-x\right)\right)}{\sinh\left(\alpha\pi\right)}-\frac{1}{2\alpha^{2}}\mathcal{H}(x,\alpha)\\
j_{E}(x,\alpha)= & \nicefrac{1}{8}\, j\left(2x,\nicefrac{\alpha}{2}\right)\\
\mathcal{H}(x\neq0,\alpha)\equiv & \sum_{k=-\infty}^{+\infty}\frac{k\sin\left(kx\right)}{\left(k-i\alpha\right)^{2}}=\frac{h(x,\alpha)+h(x,-\alpha)}{2}\\
h(x\neq0,\alpha)\equiv & -i\sum_{k=-\infty}^{+\infty}\frac{k\exp\left(ikx\right)}{\left(k-i\alpha\right)^{2}}=\frac{\pi\exp\left(-x\alpha\right)}{1-\cosh\left(2\pi\alpha\right)}\left[x\alpha\left(e^{2\pi\alpha}-1\right)+2\pi\alpha-\left(e^{2\pi\alpha}-1\right)\right]\\
k(x,\alpha)\equiv & \sum_{k=-\infty}^{+\infty}\frac{k^{3}\sin\left(kx\right)}{\left(k^{2}+\alpha^{2}\right)^{2}}=\frac{\pi}{2}\frac{\sinh\left(\alpha\left(\pi-x\right)\right)}{\sinh\left(\alpha\pi\right)}+\frac{\mathcal{H}(x,\alpha)}{2}\\
k_{E}(x,\alpha)= & \nicefrac{1}{2}\, k\left(2x,\nicefrac{\alpha}{2}\right)\\
s(x,\alpha)\equiv & \sum_{k=-\infty}^{+\infty}\frac{k^{2}\exp\left(ikx\right)}{\left(k^{2}+\alpha^{2}\right)^{2}}=\frac{\pi}{2}\frac{\cosh\left(\alpha\left(\pi-x\right)\right)}{\alpha\sinh\left(\alpha\pi\right)}+\frac{\pi}{4}u(x,\alpha)\\
s_{E}(x,\alpha)= & \nicefrac{1}{4}s(2x,\nicefrac{\alpha}{2})\\
u(x,\alpha)\equiv & \frac{2x\cosh\left(\alpha\left(x-2\pi\right)\right)+\left(2\pi-x\right)\cdot2\cosh\left(\alpha x\right)}{\left(1-\cosh\left(2\pi\alpha\right)\right)}
\end{alignat*}
The function $j_{O}$ is obtained by writing $j(x,\alpha)=j_{O}(x,\alpha)+j_{E}(x,\alpha)$;
the same applies for the other functions with subscripts \emph{O}.

The coincidence of the infinite summations and their analytical expressions
has been  {\emph{ }verified} numerically for particular
values of the parameters.

As a technical remark, we would like to mention that the preceding
formulae are difficult to evaluate numerically for $\left|\alpha\right|\gg1$,
on account of the large arguments of the hyperbolic functions. Nevertheless,
the following approximations provide very satisfactory results in
the limit of large $\alpha$'s ($\alpha>0$)~: 
\[
\frac{\sinh\left[\alpha\left(\pi-x\right)\right]}{\sinh\left(\alpha\pi\right)}\approx\exp\left(-x\alpha\right)-\exp\left(\alpha\left(x-2\pi\right)\right)
\]
 
\[
\frac{\cosh\left[\alpha\left(\pi-x\right)\right]}{\sinh\left(\alpha\pi\right)}\approx\exp\left(-x\alpha\right)+\exp\left(\alpha\left(x-2\pi\right)\right)
\]

\[
h(x,\alpha)\approx-2\pi\,\exp\left(-x\alpha\right)\left[x\alpha-1\right]
\]
 
\[
u\left(x,\alpha\right)\approx-2\left[x\,\exp\left[\alpha\left(x-4\pi\right)\right]+x\,\exp\left(-\alpha x\right)+\left(2\pi-x\right)\cdot\exp\left[\alpha\left(x-2\pi\right)\right]\right]
\]

Our final result is: 
\begin{equation}
\left(\begin{array}{c}
\sigma_{xx}^{corr}(m,n)\\
\sigma_{xy}^{corr}(m,n)
\end{array}\right)=\left(\begin{array}{c}
\frac{-2p_{m}q_{n}^{2}}{\underline{q}^{4}}\left[i{\displaystyle \sum_{y}}\zeta_{\delta}(X)\mathcal{F}_{x}\sigma_{xy}^{pl}(m,y)+2{\displaystyle \sum_{y}}\xi_{\delta}(X)\mathcal{F}_{x}\sigma_{xx}^{pl}(m,y)\right]\\
\frac{q_{n}\left(p_{m}^{2}-q_{n}^{2}\right)}{\underline{q}^{4}}\left[i{\displaystyle \sum_{y}}\zeta_{\delta}(X)\mathcal{F}_{x}\sigma_{xy}^{pl}(m,y)+2{\displaystyle \sum_{y}}\xi_{\delta}(X)\mathcal{F}_{x}\sigma_{xx}^{pl}(m,y)\right]
\end{array}\right),\label{eq:Appendix_eps_corr}
\end{equation}
where we should note that $\zeta(0,n\in O)=\frac{-2}{L_{y}^{2}}$.

As a computational detail, note that the \emph{y}-coordinates are
here integers shifted by half unity, i.e., of the form $p+\nicefrac{1}{2},\, p\in\mathbb{N}$,
whereas computational routines for Fast Fourier Transform take as
input an array with integer indices. It is therefore easier to suppose
that the walls are at positions $y=-\nicefrac{1}{2}$ and $y=L_{y}-\nicefrac{1}{2}$.
This translation is readily achieved by simply multiplying the Fourier
components of the correction term, as given above, by prefactors $\mathrm{exp}\left(\frac{iq_{n}}{2}\right)$.

Assuming a complexity $\mathcal{O}\left(N\,\ln N\right)$ for the
Fast Fourier Transform of an array of \emph{N }cells, the number of
operations performed at each time step of our algorithm is of order
$\mathcal{O}\left(L_{x}L_{y}^{2}\,\ln L_{x}\right)$ for large integers $L_{y}$
and $L_{x}$, as is evident from Eq.\ref{eq:Appendix_eps_corr}.

\section{Calculation of the line-averaged velocity\label{sec:Derivation-of-the-velocity}}

The mean velocity on a line $y=y_{0}$ reads:

\begin{eqnarray*}
\left\langle u_{x}\right\rangle _{x}\left(y_{0}\right) & \equiv & \frac{1}{L_{x}}{\displaystyle \int_{-\nicefrac{L_{x}}{2}}^{\nicefrac{L_{x}}{2}}}u_{x}\left(x,y_0\right)dx\\
 & = & \sum_{n=-\infty}^{+\infty}\hat{u}_{x}(m=0,n)e^{iq_{n}y_{0}}\\
 & = & \sum_{\underset{n\neq0}{n=-\infty}}^{+\infty}\hat{u}_{x}^{\star\infty}(0,n)e^{iq_{n}y_{0}}+\hat{u}_{x}^{\star\infty}\left(0,0\right)-\left(1-\nicefrac{2\left|y_{0}\right|}{L_{y}}\right)\sum_{I}\hat{u}_{x}^{\star\infty}(0,\cdot)+\sum_{P}\overset{0}{\overbrace{\hat{u}_{x}^{corr}(0,\cdot)}}e^{iq_{n}y_{0}}\\
 & = & \sum_{y_{ev}}\frac{a}{2\mu}\left[\mathrm{Sign}\left(y_{0}-y_{ev}\right)\cdot\left(1-\frac{\left|y_{0}-y_{ev}\right|}{L_{y}}\right)+1-\frac{y_{ev}}{L_{y}}-\frac{y_{0}}{L_{y}}\right]\mathcal{F}_{x}\sigma_{xy}^{pl}(m=0,y_{ev}),
\end{eqnarray*}

where the last summation is performed over all streamlines $y_ev$, and $\hat{u}_{x}^{\star\infty}$ is the bulk contribution in the
duplicated system.

\section{Estimation of the deviations due to bulk cooperativity\label{sec:Appendix_Babel_number}}

Assume the fluidity diffusion equation is a valid approximation, 
\[
\xi^{2}\Delta f-\left(f-f_{bulk}\right)=0
\]
where $f=\frac{\dot{\gamma}}{\sigma}$ is the local fluidity, and $\xi$
is a cooperativity length that may vary with the shear rate.

Let $\delta f=f-f_{bulk}$ be the deviation from the expected fluidity
profile owing to cooperative effects between regions subject to different
driving forces.

One now assumes $\delta f\ll f_{bulk}$ and $\Delta\delta f\ll\Delta f_{bulk}$.

To leading order, the fluidity diffusion equation reads 
\[
\xi^{2}\Delta f_{bulk}=\delta f
\]
The amplitude of the deviations due to cooperativity is given by the
Babel number $\mathrm{Ba}\equiv\frac{\delta f}{f}\approx\xi^{2}\frac{\Delta f_{bulk}}{f_{bulk}}$

If the flow curve follows a Herschel-Bulkley law: $\sigma\left(\dot{\gamma}\right)=\sigma_{d}+A\dot{\gamma}^{n}$,
{\small
\[
f_{bulk}^{\prime\prime}=\frac{\sigma^{\prime^{2}}}{A^{\nicefrac{1}{n}}}\frac{\sigma^{n-1}\left(\sigma-\sigma_{d}\right)^{\nicefrac{1}{n}-1}}{n}\left[\left(\nicefrac{1}{n}-1\right)\frac{\sigma^{-n}}{\sigma-\sigma_{d}}\left(\left(1-n\right)+n\frac{\sigma_{d}}{\sigma}\right)^{2}-n\sigma^{-n-1}\left(1-n+\left(1+n\right)\frac{\sigma_{d}}{\sigma}\right)\right]
\]
}
Here, the primes denote derivatives with respect to the space coordinate.
Then, 
\[
\frac{f_{bulk}^{\prime\prime}}{f_{bulk}}=\frac{\sigma^{\prime^{2}}}{n\left(\sigma-\sigma_{d}\right)}\left[\frac{\left(\nicefrac{1}{n}-1\right)}{\sigma-\sigma_{d}}\left(\left(1-n\right)+n\frac{\sigma_{d}}{\sigma}\right)^{2}-\text{\ensuremath{\frac{n}{\sigma}}}\left(1-n+\left(1+n\right)\frac{\sigma_{d}}{\sigma}\right)\right]
\]
To leading order, one finally arrives at $\frac{\delta f}{f}\sim\xi^{2}\frac{\sigma^{\prime^{2}}}{\left(\sigma-\sigma_{d}\right)^{2}}$. 
\end{document}